\documentclass[useAMS,usenatbib,fleqn]{mn2e}
\usepackage{courier,amssymb,amsmath,color,verbatim,graphicx,url}
\usepackage[normalem]{ulem}

\title[CFHTLenS and RCSLenS Testing Photometric Redshift
Distributions]{CFHTLenS and RCSLenS:  Testing Photometric Redshift
  Distributions Using Angular Cross-Correlations with Spectroscopic Galaxy Surveys}

\author[Choi et al.]{A.~Choi$^{1}$\thanks{E-mail: choi@roe.ac.uk},
  C.~Heymans$^{1}$,
  C.~Blake$^{2}$,
  H.~Hildebrandt$^{3}$, C.~A.~J.~Duncan$^{1}$, T.~Erben$^{3}$,
  \newauthor  R.~Nakajima$^{3}$, L.~Van~Waerbeke$^4$
  \&
  M.~Viola$^5$\\
$^{1}$Scottish Universities Physics Alliance, Institute for Astronomy,University of Edinburgh, Royal
Observatory, Blackford Hill,\\ Edinburgh EH9 3HJ, UK\\
$^2$Centre for Astrophysics \& Supercomputing, Swinburne University of Technology, PO Box 218, Hawthorn, VIC 3122, Australia\\
$^3$Argelander-Institut f\"ur Astronomie, Auf dem H\"ugel 71, 53121 Bonn, Germany\\
$^4$University of British Columbia, Department of Physics and Astronomy, 6224 Agricultural Road, Vancouver, B.C. V6T 1Z1, Canada\\
$^5$Leiden Observatory, Leiden University, Niels Bohrweg 2, 2333 CA Leiden, The Netherlands\\
}

\begin{document}

\date{\today}

\pagerange{\pageref{firstpage}--\pageref{lastpage}} \pubyear{2015}

\maketitle

\label{firstpage}

\begin{abstract}
We determine the accuracy of galaxy redshift distributions as
estimated from photometric redshift probability distributions $p(z)$.
Our method utilises measurements of the angular cross-correlation
between photometric galaxies and an overlapping sample of galaxies
with spectroscopic redshifts.  We describe the redshift leakage from a
galaxy photometric redshift bin $j$ into a spectroscopic redshift bin
$i$ using the sum of the $p(z)$ for the galaxies residing in bin $j$.
We can then predict the angular cross-correlation between photometric
and spectroscopic galaxies due to intrinsic galaxy clustering when $i
\neq j$ as a function of the measured angular cross-correlation when
$i=j$.  We also account for enhanced clustering arising from lensing
magnification using a halo model.  The comparison of this prediction
with the measured signal provides a consistency check on the validity
of using the summed $p(z)$ to determine galaxy redshift distributions
in cosmological analyses, as advocated by the Canada-France-Hawaii
Telescope Lensing Survey (CFHTLenS).  We present an analysis of the photometric redshifts measured by CFHTLenS, which overlaps the Baryon Oscillation Spectroscopic Survey (BOSS).  We also analyse the Red-sequence Cluster Lensing Survey (RCSLenS), which overlaps both BOSS and the WiggleZ Dark Energy Survey.  We find that the summed $p(z)$ from both surveys are generally biased with respect to the true underlying distributions.  If unaccounted for, this bias would lead to errors in cosmological parameter estimation from CFHTLenS by less than $\sim 4\%$.  For photometric redshift bins which spatially overlap in 3-D with our spectroscopic sample, we determine redshift bias corrections which can be used in future cosmological analyses that rely on accurate galaxy redshift distributions.

\end{abstract}


\section{Introduction}
\label{sec:intro}

Cosmological parameter estimation often relies on highly accurate
knowledge of the underlying 3-D spatial distributions of the galaxies
used in the analysis.  The most direct way to estimate the distributions in the redshift
dimension is to measure the redshifts of all galaxies of interest using high
resolution information from spectroscopy, but this is
not only costly but potentially incomplete due to the difficulties of
measuring secure redshifts for certain populations of galaxies \citep{2014MNRAS.444..129C,masters/etal:2015}.  Photometric redshift estimation provides a lower resolution and
less expensive tool for constraining redshift distributions.  While
fundamentally limited by the available filters, relevant model and assumptions (such as template
choice or how representative the training set is), summed redshift probability distribution functions, $p(z)$, provide estimates of the underlying redshift distributions that are less biased than counts of single-point
estimates when compared against spectroscopic information \citep{2008MNRAS.386..781M,2009MNRAS.396.2379C,2009ApJ...700L.174W,2011ApJ...734...36A,2012MNRAS.420.3240N,2012ApJS..201...32S,2013MNRAS.431.1547B}.

Various techniques for calibration of these errors and the distribution of photometric galaxies have
 been promoted \citep[for a review see][]{2015APh....63...81N}, falling roughly into three categories:
\begin{itemize}
\item \textit{Direct
   calibration} requires a complete and representative training
 sample with which to re-weight the photometric galaxies or compare
 them on an individual basis
 \citep{2012MNRAS.421.1671B,2015arXiv150705909B}.  The spectroscopic data
sets that are used to characterise photometric redshift
scatter and bias are often only complete to a magnitude that is much
brighter than the magnitude of a typical galaxy used in cosmological analyses.
\item \textit{Reconstruction} methods utilise the fact that
there is an excess probability of pairs of galaxies (relative to a
Poisson distribution) that are truly physically correlated in 3-D.
The clustering information is then used to infer the true underlying
redshift distributions \citep{2006ApJ...651...14S,2008ApJ...684...88N,2010ApJ...721..456M,2010ApJ...724.1305S,2013MNRAS.431.3307S,2013MNRAS.433.2857M,2013arXiv1303.4722M,2014ApJ...780..185D,2015MNRAS.447.3500R}.
This strategy needs a spectroscopic sample to span the
full redshift range that can be incomplete in terms of galaxy properties. Cross-correlations measure the combination of the galaxy bias times the redshift probability distribution, and thus all of these methods require additional constraints on the galaxy bias of the given sample to break the degeneracy.  \citet{2008ApJ...684...88N} propose an iterative procedure to account for evolution in the galaxy bias based on the auto-correlations of the spectroscopic and photometric samples, respectively, although the effectiveness of the correction depends on the shape of the redshift distribution as well as the linearity of the galaxy bias evolution  \citep[for further discussion see][]{2008ApJ...684...88N,2013MNRAS.431.3307S}.  In principle, the
cross-correlation strategies have a key advantage over standard direct
calibration methods as they do not rely as strongly on the
completeness of the sample with available spectroscopic redshifts
provided that the spectroscopic redshifts cover a 3-D space
overlapping with the photometric redshifts.
\item \textit{Verification} methods use the cross-clustering signal between galaxies in different redshift
bins to indicate the degree of contamination between
those redshift bins and test for consistency with other
estimates of the redshift distribution such as summed $p(z)$.
\citet{2009A&A...493.1197E,2010MNRAS.408.1168B,2013MNRAS.431.1547B}
investigated the photometric-photometric case, and here we extend the
formalism to photometric-spectroscopic samples.  The advantage of this approach over others is that it can yield
constraints on catastrophic outliers even when the spectroscopic
sample does not extend over the full redshift range under consideration.
\end{itemize}

There are three types of photometric
redshift error.  Two of these, random scatter and systematic bias, move galaxies into
adjacent or closely neighbouring redshift bins.  A third kind occurs when systematic bias sends galaxies into distant redshift bins and is commonly referred to as a `catastrophic
outlier'.  With the angular cross-correlation analyses used in reconstruction and verification methods, a detection of a strong clustering signal
between a low redshift spectroscopic sample and a high redshift
photometric sample is an indication that catastrophic outliers
exist in the photometric sample.  However, one must also consider
astrophysical effects caused by lensing magnification, which most
previous works have ignored.  Magnification can be thought of as both a 
contaminant to redshift recovery via cross-correlations 
\citep[][]{2010MNRAS.401.1399B} and an informative 
signal in its own right \citep[][]{2005ApJ...633..589S,2009A&A...507..683H,2010MNRAS.405.1025M,2012MNRAS.426.2489M,2014MNRAS.437.2471D}, 
containing the imprint of galaxy evolution and 
cosmological processes.  Lensing of light by foreground structures 
(de)magnifies images of background galaxies.  Hence, at a fixed apparent magnitude the 
number density behind a massive foreground galaxy will change, which will be 
seen in an angular cross-correlation signal.  A cross-correlation signal can contain contributions from both magnification and catastrophic outliers, thus necessitating a careful investigation of the magnification effects before a clear interpretation can be made about the presence of catastrophic outliers.  \citet{1998MNRAS.294L..18M} present the theory behind the lensing and intrinsic contributions to the 
total observed clustering signal as well as investigate the cosmological 
dependence.

In this paper, we outline formalism for testing consistency between the
estimated redshift probability distributions of a photometric sample and angular cross-correlations
between the photometric sample and a spatially overlapping
spectroscopic sample (Section 2).  Our verification method fully
accounts for the unknown galaxy bias, assuming that the average galaxy
bias of an outlier population at a given photometric redshift does not significantly deviate from the
average galaxy bias of the main population at the same photometric redshift (see Section 2.3 for the details).  This
is a notable advantage compared with the aforementioned clustering-based reconstruction
approaches for which the redshift probability distribution is
completely degenerate with galaxy bias (before additional corrections).  We model the effects of the magnification 
component using the halo model.  We then apply this test to a $\sim 66$ deg$^{2}$ region where there are spatially overlapping samples of photometric galaxies imaged by the Canada-France-Hawaii Telescope Lensing Survey (CFHTLenS) and galaxies with spectroscopic redshifts from the BOSS survey and a $\sim 200$ deg$^{2}$
 region with overlap between the Red Sequence Cluster Lensing Survey (RCSLenS) and
 both the WiggleZ and BOSS surveys.  The considerable amount of
 spectroscopic overlap makes CFHTLenS and RCSLenS ideal data sets on which to
 test cross-correlation techniques.  We describe these surveys and the catalogue production pipeline in Section 3.  In Section 4, we present
the angular cross-correlation measurements and predictions
based on our models for the intrinsic and magnification clustering and discuss the level
of consistency.  We conclude in Section 5.  In the Appendix, we
describe validation tests of our method on mock galaxy catalogues, provide further details of the halo modelling used, and check for systematics in the object catalogues.

For the modelling of the magnification signal only, we assume cosmological parameters from \citet{2014A&A...571A..16P}, with $\Omega_{\rm m}=0.315$, $\Omega_{\Lambda}=0.685$, $\sigma_{8}=0.829$, $n_{\rm s}=0.9603$ and $\Omega_{\rm b}h^{2}=0.02205$.

\section{Formalism}

In this section, we present the formalism for the angular
cross-correlations that are the focus of this work, beginning with a
general discussion of galaxy clustering and the physical processes
that contribute to the signal.  We then relate the redshift
probability distributions to
the intrinsic angular cross-correlations between spectroscopic and photometric
samples and describe how we use the halo model to construct predictions for the 
angular cross-correlations arising from lensing magnification.

\subsection{Angular Correlations and Magnification}

The clustering signal $w(\theta)$ is a two-point angular correlation
function, or the excess probability of finding a pair of objects in a
solid angle $d\Omega$ and angular separation $\theta$ such that the
probability element is given by
\begin{equation}
dP = N[1+w(\theta)]d\Omega\, ,
\end{equation}
where N is the mean number of galaxies per unit steradian \citep{1973ApJ...185..413P}. For two different 
galaxy samples at mean redshifts $\langle z_{1} \rangle$ and 
$\langle z_{2} \rangle$, 
\begin{equation}
dP_{1,2} = N_{1}N_{2}[1+w_{1,2}(\theta)]d\Omega_{1}d\Omega_{2}\, .
\end{equation}

In practice, we use the Landy-Szalay estimator \citep{LandySzalay93}
for $w_{1,2}(\theta)$ given by,

\begin{equation}\label{eq:LS}
  \begin{split}
    w_{1,2}(\theta) = \frac{(D_{1}D_{2})_{\theta}}{(R_{1}R_{2})_{\theta}}
    \frac{N_{R,1}N_{R,2}}{N_{1}N_{2}} -
    \frac{(D_{1}R_{2})_{\theta}}{(R_{1}R_{2})_{\theta}}
    \frac{N_{R,2}}{N_{1}} \\
    - \frac{(D_{2}R_{1})_{\theta}}{(R_{1}R_{2})_{\theta}}
    \frac{N_{R,1}}{N_{2}} + 1 \, .
  \end{split}
\end{equation}
$(D_{1}D_{2})_{\theta}$ is the number of pairs with one galaxy in data
sample $1$ and the other in data sample $2$ as a function of the
angular separation $\theta$.  Similarly, $(D_{1}R_{2})_{\theta}$ is
the number of pairs with one galaxy in data sample $1$ and the other
in a random sample $2$, which is constructed to reflect the same
selection properties like masks and geometry as the corresponding data
sample $2$. $(R_{1}R_{2})_{\theta}$ is the number of pairs with one
galaxy in random sample $1$ and the other in random sample $2$.  It is
necessary to use random catalogues that are many times more highly sampled than the data catalogues in order to minimise the noise contributed by including the additional random pair counts.  Therefore, each of the terms in Eqn.~\ref{eq:LS} must be normalised by factors involving $N_{1}$, $N_{2}$, $N_{R,1}$, $N_{R,2}$ representing the number of galaxies in data sample $1$, the number in data sample $2$, the number in random sample $1$, and the number in random sample $2$, respectively. 

We can also write 
$w_{1,2}(\theta)$ in terms of differential number densities
\begin{equation}\label{eq:wtheta_deltan}
  w_{1,2}(\theta) = \langle \delta n_{1}(\phi) \delta n_{2}(\phi + \theta) \rangle_{\phi} \, ,
\end{equation}
where
\begin{equation}\label{eq:deltan}
  \delta n_{i}(\phi) \equiv \frac{n_{i}(\phi)-\bar{n}_{i}}{\bar{n}_{i}}
  = \delta n_{i}^{g}(\phi) + \delta n_{i}^{\mu}(\phi)\, ,
\end{equation}
and $n_{i}(\phi)$ is the number density of galaxies belonging to a 
sample at redshift $\langle z_{i} \rangle$ observed at position angle $\phi$, and 
$\bar{n}_{i}$ is the average density of the $i$th sample. The superscript $g$ 
indicates the intrinsic galaxy component and the 
superscript $\mu$ indicates the magnification component.  Combining Eqn.~\ref{eq:deltan} and Eqn.~\ref{eq:wtheta_deltan} yields
\begin{equation}\label{eq:wtheta_alldelta}
  \begin{split}
  w_{1,2}(\theta) = \langle \delta n_{1}^{g}(\phi) \delta n_{2}^{g}(\phi+\theta) \rangle_{\phi} + \langle \delta n_{1}^{g}(\phi) \delta n_{2}^{\mu}(\phi+\theta) \rangle_{\phi} \\ + 
\langle \delta n_{1}^{\mu}(\phi) \delta n_{2}^{g}(\phi+\theta) \rangle_{\phi} + \langle \delta n_{1}^{\mu}(\phi) \delta n_{2}^{\mu}(\phi+\theta) \rangle_{\phi} \, .
  \end{split}
\end{equation}
$w_{1,2}(\theta)$ contains 
four terms.  The first is the intrinsic galaxy clustering due to 
gravity.  In the limit that sample 1 and sample 2 do not overlap, this term disappears.  Any redshift overlap between samples 1 
and 2 increases the strength of this term.
The second and third components arise from the lensing magnification and depend on the amount of overlap.  $w_{1,2}(\theta)$ is dominated by the second term 
when $\langle z_{1} \rangle < \langle z_{2} \rangle$ and there is no redshift overlap between samples 1 and 2.  
The fourth term is due to pure matter-matter correlations, which we will
ignore for the remainder of this work as it is sub-dominant to 
the other terms in every case considered here \citep{2011MNRAS.415.1681H,2014MNRAS.437.2471D}.

Focusing on the first two terms, respectively, and assuming linear bias:
\begin{equation}
  \label{eq:intrinsiccl}
  \begin{split}
  \langle \delta n_{1}^{g}(\phi) \delta n_{2}^{g}(\phi+\theta) \rangle_{\phi} &= b_{1}b_{2} \int_{0}^{\chi_{H}} d\chi \,\eta_{1}(\chi) \eta_{2}(\chi) \\
  &\times \int_{0}^{\infty} \frac{k dk}{2\pi} \,P_{\delta}(k, \chi) J_{0}(\chi k \theta)\, ,
  \end{split}
\end{equation}
where $b_{i}$ is the galaxy bias of sample $i$, $\chi$ is the co-moving distance,  P$_{\delta}$ is the 3-D dark matter
power spectrum, $\eta_{i}$ is the co-moving distance distribution of
sample $i$ and $J_{0}$ is the zeroth
order Bessel function.
\begin{equation}
  \label{eq:magcl}
  \begin{split}
    \langle \delta n_{1}^{g}(\phi) \delta n_{2}^{\mu}(\phi+\theta) \rangle_{\phi} &= b_{1}(\alpha - 1) \int_{0}^{\chi{_H}} d\chi \eta_{1}(\chi) K(\chi) \\
    &\times \int_{0}^{\infty} \frac{kdk}{2\pi} P_{\delta}(k, \chi)J_{0}(\chi k \theta)\, ,
  \end{split}
\end{equation}
where $\alpha$ is the slope
of the magnitude number counts, defined formally as
\begin{equation}\label{eq:magalpha}
\alpha(r) = 2.5 \frac{d\log_{10} n(>r)}{dr}\, ,
\end{equation}
with $n$ the observed galaxy number density and $r$ the galaxy magnitude.  $\alpha$ is measured using the appropriate detection band ($r$-band for RCSLenS and $i$-band for CFHTLenS).
$K(\chi)$ is the lensing kernel-weighted
distribution of background sources $\eta_{2}(\chi)$ defined as,
\begin{equation}
  K(\chi) = \frac{3H_{0}^{2}\Omega_{m}}{c^{2}} \frac{\chi}{a}
    \int_{\chi}^{\chi_{H}} d\chi^{\prime} \eta_{2}(\chi^{\prime})
    \frac{\chi^{\prime} - \chi}{\chi^{\prime}}\, ,
\end{equation}
with $H_{0}$ the Hubble constant today, $a$ the scale factor and assuming a flat universe.

Magnification affects the clustering
in two ways.  First, it raises the flux of a magnified galaxy such that the
galaxy count might be boosted within the flux limit of a survey.  Second, it
increases the observed solid angle around a magnifying galaxy.  The net effect, given by Eqn.~\ref{eq:magcl}\footnote{The magnification contribution to the clustering signal of off-diagonal redshift bin combinations can be positive
or negative.}, depends on the slope of the luminosity function of the background
objects, $\alpha$, defined in Eqn.~\ref{eq:magalpha}.  See \citet{2005ApJ...633..589S,2009A&A...507..683H,2010MNRAS.405.1025M,2012MNRAS.426.2489M} for measurements of the magnification signal via number counts.
We measure $\alpha$ for
RCSLenS and CFHTLenS by calculating the slope of the cumulative number densities of galaxies as a function of limiting magnitude and binned by photometric redshift \citep{2014MNRAS.437.2471D}.

We use the halo model
implemented in the \textsc{Python} package
\textsc{CHOMP}\footnote{http://code/google.com/p/chomp} to generate
theoretical predictions for the magnification contribution to $w(\theta)$ for BOSS galaxies.  We take the BOSS halo occupation distribution (HOD) parameters determined in 
\citet{2013MNRAS.429...98P} for the LOWZ sample and
\citet{2013arXiv1311.1480M} for the CMASS sample. We do not estimate
predictions for the magnification around WiggleZ galaxies due to the
difficulty of obtaining an HOD description of this sample, which is
not volume-limited.  The WiggleZ galaxies have lower masses, so we expect the amplitude of
the signal to be negligible in the one-halo regime.  The signal might
be more comparable in the regime
where the two-halo term dominates (roughly the largest 2-3 $\theta$
bins), but the S/N of the data over these last few bins does not
warrant the modelling as we will see in Section 4.3.  Appendix~\ref{sec:halomodel} contains
further details of the halo model and the input HOD parameters to the estimated magnification signal.

\subsection{Contamination From Photometric Redshift Errors}
\label{sec:methodology}

In this section, we extend the formalism presented in
\citet{2010MNRAS.408.1168B,2013MNRAS.431.1547B}, who considered
 cross-correlations
between photometric redshift bins.  Here, we examine the case of cross-correlations between spectroscopic and photometric redshift bins.  We define the following quantities:

\begin{itemize}
  \item Each galaxy has a single-point best fitting photometric redshift z$_{B}$ and 
    photometric redshift probability distribution $p(z)$.
  \item $N_{j}^{\rm O}$ is the observed number of galaxies with
    single-point photometric redshifts that place them in bin $j$.
  \item   When discussing aggregate or summed $p(z)$, we use the notation $\Phi_{j}(z)$ as the sum of the $p(z)$ for all galaxies in photometric redshift bin $j$,
\begin{equation}
  \label{eq:pz}
  \Phi_{j}(z)=\sum_{k=1}^{N_{j}^{O}}p_{kj}(z)\, .
\end{equation}
  \item $N_{ij}^{\rm T}$ is the true number of galaxies which would have a spectroscopic redshift in bin
    $i$ (if spectroscopy had been measured for those galaxies) and placed in photometric redshift bin $j$.  $N_{ii}^{\rm T}$ is
    then the number of galaxies that have both spectroscopic and photometric redshifts placing them in bin $i$.
  \item $w_{ij}^{\rm SP,O}$ is the observed clustering signal between
    galaxies with photometric redshifts in bin $j$ and galaxies in the spectroscopic sample with
    spectroscopic redshifts in bin $i$ and includes all terms from Eqn.~\ref{eq:wtheta_alldelta}.
  \item $w_{ij}^{\rm SP,T}$ is the true clustering signal between
    galaxies in the full photometric sample with spectroscopic redshifts in bin $j$ and those in the spectroscopic sample with spectroscopic redshifts in bin $i$.
\end{itemize}

If $i \neq j$ (which we will often refer to as ``off-diagonal''), in the absence of lensing magnification, the true cross-clustering signal
$w_{ij}^{\rm SP,T}=0$.  Consider the pairwise case of the first two bins.
When $i=j$ and the spectroscopic and photometric redshift bins are specified to be the same
range (``diagonal''), the observed cross-clustering signal is related 
to the true
cross-clustering signal. The true signal is scaled by the ratio of the true number of galaxies
assigned to photometric bin $j=1$, which actually have a true redshift in
the same spectroscopic bin $i=1$, to the total observed
number of galaxies in photometric bin $j=1$.  When the spectroscopic and photometric redshift 
bins are different (off-diagonal, e.g. $i=1$ and $j=2$) but still overlapping due to photometric redshift scatter and outliers, the observed cross-clustering signal will be the
true diagonal cross-clustering signal multiplied by the ratio of the true
number of galaxies assigned to photometric bin $j=2$, which actually have a
true redshift in spectroscopic bin $i=1$, to the total
observed number of galaxies in photometric bin $j=2$.  Written using the quantities
defined above and generalizing to the case of multiple bins\footnote{Eqn. 4 from \citet{2013MNRAS.431.1547B} describes the observed
cross-clustering signal between two photometric redshift bins $i$ and $j$.  In this analysis, one bin is always spectroscopic, so that the fractional leak from the spectroscopic bin 1 to the photometric bin 2 is always zero.  In this case, the formalism presented here is consistent with \citet{2013MNRAS.431.1547B}.},

\begin{eqnarray}\label{eq_crossterm}
  w_{ij}^{\rm SP,O} &=& \frac{w_{ii}^{\rm SP,T}N_{ij}^{\rm T}}{N_{j}^{\rm O}} \, .
\end{eqnarray}

\subsection{Procedure for Estimating the Clustering Signal Using $\Phi_{j}(z)$}
\label{sec:procedure}

We outline the steps to model the intrinsic clustering signal
(i.e. Eqn.~\ref{eq:intrinsiccl}) of off-diagonal
redshift bin combinations using photometric redshift error distributions.  We use E($x$) to denote the estimator for the quantity $x$ defined in Section~\ref{sec:methodology}.

\begin{itemize}
\item \textit{Step 0:} Create spectroscopic and photometric samples using redshift bin limits based on available spectra.  In this work, we use the same bin limits for both spectroscopic and photometric bins due to signal to noise considerations.  In principle, however, the spectroscopic information has much higher resolution and can potentially be binned more finely.
\item \textit{Step 1:} Measure the observed number of galaxies in each photometric bin $N_{j}^{\rm O}$ by counting the single-point
 photometric redshift estimates $z_{B}$.
\item \textit{Step 2:} Measure the auto- and cross-correlations between spectroscopic and photometric bins and corresponding covariance
matrices.  In this work, we use jack-knife resampling to estimate the covariance matrices. 
\item \textit{Step 3:} Estimate E($N_{ij}^{\rm T}$) by summing the
  $p(z)$ corresponding to galaxies selected to be in a photometric redshift bin $j$
using $z_{B}$ to obtain $\Phi_{j}(z)$ (Eqn.~\ref{eq:pz}) and then
integrating $\Phi_{j}(z)$ over the limits of true-z bin
$i$.  Note that E(N$_{ij}^{\rm T}$) summed over all bins $i$ must equal
$N_{j}^{\rm O}$.  In practice, we implement this constraint by ensuring
that $\Phi_{j}(z)$ integrated from the minimum to the maximum of
the full redshift range normalises to $N_{j}^{\rm O}$.

\item \textit{Step 4:} Use the E(N$_{ij}^{\rm T}$) to predict the
  observed cross-correlation, which can be derived from Eqn.~\ref{eq_crossterm} as
\begin{equation}
\label{eqn:Ewijmodel}
  {\rm E}(w_{ij}^{\rm SP,O})= w_{ii}^{\rm SP,O} \frac{N_{i}^{\rm O}}{N_{j}^{\rm O}}E(\frac{N_{ij}^{\rm T}}{N_{ii}^{\rm T}}) \, .
\end{equation}

\item \textit{Step 5:} Compare the predicted E($w_{ij}^{\rm SP,O}$) and observed cross-correlation $w_{ij}^{\rm SP,O}$.  If they agree, $\Phi_{j}(z)$ is a good estimate of the redshift distribution of the galaxy sample.
\end{itemize}
We present a validation of this method on an idealised mock galaxy catalogue in Appendix~\ref{sec:valmocks}.

By constructing the estimate in this way, we bypass having to constrain the true galaxy bias of the spectroscopic sample in each bin, $b_{i}$, and the galaxy bias of our photometric sample in each bin, $b_{j}$, since they are included within the measurement of the
diagonal cross-correlations as can be seen in Eqn.~\ref{eq:intrinsiccl}.  However, we must assume that $b_{j}$ has
either no or slow evolution over the full extent of the distribution in each photometric bin and, more importantly, that catastrophic outliers are a random sample of the bin's galaxy population such that the average galaxy bias of the outliers is the same as the average bias of the whole population.  We revisit these assumptions in Section~\ref{sec:galbias}.

\subsection{Goodness of fit}

To compare the predicted model and observed cross-correlation signal we calculate a goodness of fit as measured by
$\chi^{2}$.  We define the 
$\chi^{2}$ of the fit of the model prediction to the data as
\begin{equation}\label{eq:chisq}
  \chi^{2} = (\boldsymbol{d}-\boldsymbol{m})^{T} \boldsymbol{C}^{-1}(\boldsymbol{d}-\boldsymbol{m})\, ,
\end{equation}
where $\boldsymbol{d}$ is a vector of length $N_{z} \times  N_{\theta}$ containing the measured $w_{ij}(\theta_{q})$ with $i=\left\{ 1,\cdots,N_{s} \right\}$, $j$=$\left\{1,\cdots,N_{p}\right\}$ and $q$=$\left\{ 1,\cdots,N_{\theta}\right\}$.  $N_{s}$ and $N_{p}$ are the number of spectroscopic and photometric redshift bins, respectively.  $N_{z}=N_{s} \times N_{p}-\min(N_{s},N_{p})$ is the total number of spectroscopic-photometric redshift bin combinations with $i \neq j$.  $N_{\theta}$ is the total number of angular scales for which $w_{ij}(\theta_{q})$ is calculated.  $\boldsymbol{m}$ contains the model for $w_{ij}(\theta_{q})$ as given by Eqn.~\ref{eqn:Ewijmodel}.  This model depends on $w_{ii}(\theta_{q})$, which has an associated error.  We assume the two terms $w_{ij}(\theta_{q})$ and $w_{ii}(\theta_{q})$ are independent and add the corresponding covariance matrices such that the total covariance matrix $\boldsymbol{C_{ij}}= \textrm{Cov}\left[ w_{ij}(\theta_{q}), w_{ij}(\theta_{r})\right] + f_{ij}^{2}\textrm{Cov}\left[ w_{ii}(\theta_{q}), w_{ii}(\theta_{r})\right]$ is the propagated jack-knife resampled covariance between $\theta$ scales $q$ and $r$, and $f_{ij}=(N_{i}^{\rm O}/N_{j}^{\rm O})E(N_{ij}^{\rm T}/N_{ii}^{\rm T})$ is the prefactor in Eqn.~\ref{eqn:Ewijmodel}.  Note that the spectroscopic bins are sufficiently broad such that radial correlation between bins, and thus the covariance between different spectroscopic $i$ bins in this analysis, is negligible. Throughout this work, we use angular scales in the range $1^{\prime} < \theta < 35^{\prime}$, where the limits are chosen to mitigate the impact of scale-dependent galaxy biases while still utilising the signal at intermediate scales.

\subsection{Modifying photometric redshift distributions}
\label{sec:MCMCshift}

As we will see in Section~\ref{sec:results}, the models predicted from the $\Phi_{j}(z)$ are often a poor fit to the data.  We therefore investigate two methods to modify the redshift distributions for galaxy samples binned by photometric redshift.

The first modification takes the $\Phi_{j}(z)$ and allows for a shift
along the $z$ dimension in the overall distribution.   This represents
a correction for a bias error in the measured $\Phi_{j}(z)$, whilst
maintaining the level of scatter and catastrophic outliers as
specified by the $\Phi_{j}(z)$.   This is modelled using one free parameter per photometric redshift bin $j$, $\Delta z_{j}$.  When the probability gets shifted to negative redshifts, we re-normalise $\Phi_{j}(z)$ by the integrated $\Phi_{j}(z<0)$.  This shifting approach is similar to that adopted by the cosmological tomographic shear analysis of \citet{2015arXiv150705552T}, who allow for an independent shift of the estimated photometric redshift distribution as a nuisance parameter.

The second modification  models the redshift
distributions in each photometric redshift bin as a Gaussian.  This
model has two free parameters per photometric redshift bin $j$ with a
mean $\mu_{zj}$ and standard deviation $\sigma_{zj}$ and are defined on a baseline redshift range with $z>0$.  This test
allows us to directly determine the photometric redshift bias and
scatter in each bin, independent of the \textsc{BPZ} $\Phi_{j}(z)$.  The limitation of this single Gaussian model definition, however, is that it sets all catastrophic outliers to zero.

We determine fits for these free parameters for photometric redshift bins by jointly fitting the data using Monte Carlo Markov Chain (MCMC) sampling, as implemented with the \textsc{Python} code \textsc{emcee}\footnote{http://dan.iel.fm/emcee/current} \citep{2013PASP..125..306F}, which is an implementation of the affine invariant methods by \citet{GoodmanWeare2010}.  For a given combination of spectroscopic/photometric redshift surveys, we minimise the negative log likelihood calculated jointly for every $ij$ cross-correlation as a function of either the additive shifts to the $\Phi_{j}(z)$, or the Gaussian model parameters. In the MCMC analysis of the data, we do not include the effects of magnification as in most cases, the contribution is at the percent level on the smallest scales used (1$^{\prime}$).  

In Appendix~\ref{sec:valmocks}, we demonstrate that our methodology to determine the redshift distribution offsets is valid for the idealised case of Gaussian errors in the photometric redshift distributions.

\section{Data}

In this section, we provide details about the photometric and spectroscopic data sets we use in this study and an investigation into the dependence of photometric redshift errors on galaxy type.

\subsection{Photometric Surveys}
\label{sec:photdata}

We utilise two deep, wide, and high-resolution photometric surveys observed by
MegaCam on the 3.6-m Canada-France-Hawaii Telescope (CFHT).  The
 Canada-France-Hawaii Telescope Lensing Survey \citep[CFHTLenS;
][]{2012MNRAS.427..146H,2013MNRAS.433.2545E}\footnote{http://www.cfhtlens.org} is based on 154 deg$^{2}$
of $ugriz$ imaging
from the wide
component of the CFHT Legacy Survey.
\citet{2012MNRAS.427..146H} provide an overview of the pipeline,
and details about the data analysis can be found in the following: \citet{2013MNRAS.433.2545E}
describes the data reduction with automated
masking;  the photometry was Gaussianised to homogenise the point spread functions among the
different filters, object catalogues
were created with SExtractor \citep{BertinArnouts,2013MNRAS.433.2545E}, and
photometric redshifts were estimated using \textsc{BPZ}  \citep{2000ApJ...536..571B,2012MNRAS.421.2355H}; galaxy shapes were
estimated using Bayesian model fitting with \emph{Lens}fit
\citep{2013MNRAS.429.2858M}.   The mean seeing is 0.72$\arcsec$ (r-band) and
0.68$\arcsec$ (i-band), and the median redshift is $z_{m}=0.7$.

The Red Sequence Cluster Lensing Survey
\citep[RCSLenS; ][]{2016MNRAS.tmp.1132H}\footnote{http://www.rcslens.org}
is based on the second Red-sequence Cluster Survey (RCS2), comprising
nearly 800 deg$^{2}$ of multi-colour imaging 1-2
magnitudes deeper than SDSS \citep{2011AJ....141...94G}.  The
resolution is lensing-quality with a median seeing in the r-band of
0.7$\arcsec$.  A total of 513
deg$^{2}$ is imaged in multiple bands with $griz$, allowing photometric redshifts and $p(z)$ to be estimated.  The images were processed and the object
catalogues were created with the
same methods applied
to CFHTLenS \citep{2016MNRAS.tmp.1132H}.  
The number distribution of magnitudes as a
function of un-weighted objects led to choices of magnitude cut-offs of $i=24.7$ for CFHTLenS sources and
$r=23.7$ for RCSLenS sources.  These numbers correspond to the
5-$\sigma$ detection limit in a 2$\farcs$0 aperture \citep{2013MNRAS.433.2545E}.

\begin{figure}
  \includegraphics[width=0.495\textwidth]{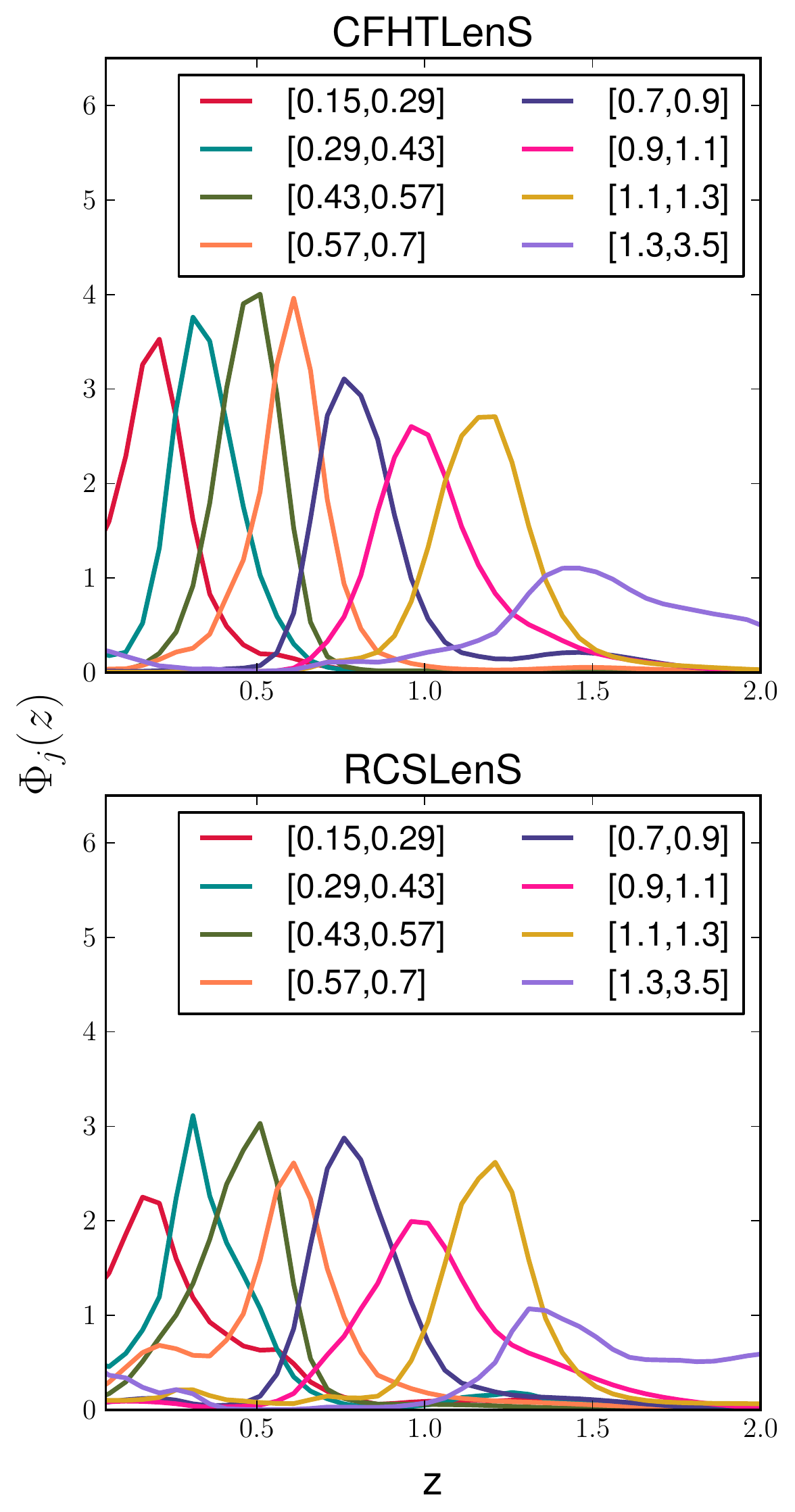}
  \caption{\label{fig:sumpdf}
   Summed probability redshift distributions $\Phi_{j}(z)$ for galaxies with single-point photometric
   redshifts $z_{B}$ in
   different bins $j$.  The top panel
   corresponds to the 66 square degrees of unmasked overlap between CFHTLenS
   and BOSS DR10.  The bottom panel corresponds to the 184 square
   degrees where there is unmasked overlap between RCSLenS and BOSS DR10.}
\end{figure}

Figure~\ref{fig:sumpdf} shows the normalised $\Phi_{j}(z)$ corresponding
to the eight photometric redshift bins considered in this analysis
with limits given by:
[0.15, 0.29],
[0.29, 0.43], [0.43, 0.57], [0.57, 0.7], [0.7, 0.9], [0.9, 1.1], [1.1,
1.3], [1.3, 3.5].  The choice of bin edges was motivated by the
spectroscopic samples used (i.e. using the LOWZ and CMASS partitions
described in Section~\ref{sec:specsurvey}), and $z=1.3$ was the
cut-off redshift for previous CFHTLenS lensing analyses, motivated by
the lack of near-infrared photometry to constrain higher redshifts.
The top panel shows the $\Phi_{j}(z)$ for the entire CFHTLenS area overlapping with BOSS DR10, while the lower panel
shows the $\Phi_{j}(z)$ for the RCSLenS overlap with BOSS DR10 and WiggleZ.  Compared with CFHTLenS, the RCSLenS $\Phi_{j}(z)$ are
noticeably multi-modal,
with flatter tails extending to adjacent bins, which reflects poorer
photometric redshift accuracy due to the limited filter coverage, most critically the missing $u$ band.

As the calculation of $w(\theta)$ requires pair counts with random
positions (see Eqn.~\ref{eq:LS}), we generate random catalogues for each field, taking into account
edges and masks.

\subsection{Spectroscopic Surveys}
\label{sec:specsurvey}

We briefly summarise the spectroscopic data sets used in this analysis but refer the interested reader to \citet{2015arXiv150703086B} for further details and statistics regarding the WiggleZ-RCSLenS, BOSS-CFHTLenS, BOSS-RCSLenS overlap regions, as they use nearly identical WiggleZ and BOSS spectroscopic samples.

\subsubsection*{BOSS}

The Baryon Oscillation Spectroscopic Survey \citep[BOSS;][]{2011AJ....142...72E} is a spectroscopic survey of massive galaxies and quasars
selected from SDSS photometry carried out at the Sloan Telescope at Apache Point Observatory in Sunspot, New Mexico, USA.  Data Release 10 (DR10) contains BOSS spectra
taken through July 2012 and comprises 927,844
galaxy spectra over 6373.2 square degrees (Ahn et al. 2013).  BOSS
galaxies were selected using colour and magnitude cuts and are divided
into the ``LOWZ'' sample with red galaxies $z<0.43$ and the ``CMASS''
sample which is designed to be approximately stellar mass-limited for
$z>0.43$.  There is a total of 66.3 deg$^{2}$ of total unmasked
overlap with CFHTLenS (W1 and W4, 2830 LOWZ galaxies, 5567 CMASS
galaxies) and a total of 183.9 deg$^{2}$ of total unmasked overlap
with RCSLenS (6 fields, 9214 LOWZ galaxies, and 18,156 CMASS
galaxies).  See Table 1 of \citet{2015arXiv150703086B} for numbers
corresponding to each field.  The 6 RCSLenS fields are labelled as
0047, 0133, 1514, 1645, 2143 and 2329.  The field named 1303 has a
very small number of galaxies with spectroscopic redshifts, and we
exclude it from this analysis.  We trim the catalogues to 
restrict them to the overlap regions (i.e. no BOSS or CFHTLenS/RCSLenS galaxies falling 
outside of the overlap are included in the analysis).

BOSS galaxies are assigned completeness weights as in Eqn. 18 of \citet{2014MNRAS.441...24A} in order to correct for the effects of redshift failures, fibre collisions and other known systematics, and we use these weights in our determination of the cross-correlations of the galaxies.  Specifically, the pair counts in Eqn.~\ref{eq:LS} are weighted by the completeness weights.

\subsubsection*{WiggleZ}

The WiggleZ Dark Energy Survey is a spectroscopic survey of bright 
emission-line galaxies with median redshift $\sim 0.6$ carried out at
the Anglo-Australian Telescope in Siding Spring, Australia
\citep{2010MNRAS.401.1429D}.  WiggleZ galaxies were selected using
colour and magnitude cuts from a combination of optical and UV
imaging.  There is a total of 175.1 deg$^{2}$ of total unmasked
overlap with RCSLenS imaging that has 4 bands and photometric redshift
estimates.  76,900 galaxies covering the range $0.1<z_{\rm s}<0.9$
reside in this region, which is comprised of 5 different RCSLenS
fields \citep[0047, 0310, 2143, 2329, 2338; for numbers corresponding to each field see Table 1 of][]{2015arXiv150703086B}.  Note that we use WiggleZ galaxies at higher
redshifts compared to \citet{2015arXiv150703086B}, who cut their
samples to $z_{\rm s}<0.7$ to match BOSS. 1514 contains a small number of galaxies with spectroscopic redshifts, and the geometry of the overlap with the RCSLenS data is irregular and patchy (the pointings are
non-contiguous). The estimated covariance matrices are not
positive-definite, and we exclude 1514 in this analysis.  We again trim the catalogues to 
restrict them to the overlap regions (i.e. no WiggleZ or RCSLenS galaxies falling 
outside of the overlap are included in the analysis).

\subsubsection*{VVDS, VIPERS and DEEP2 EGS}
\label{sec:VVD}
CFHTLenS overlaps with three small-area, but densely sampled deep spectroscopic surveys.  There are $\sim 2.6$ square degrees of spectroscopic overlap with the VIMOS VLT Deep Survey \citep[VVDS;][]{2005A&A...439..845L}.  VVDS selects objects with $17.5 \leq i \leq 24$ and is $\sim$90\% complete down to $i<23$.   There are $\sim 0.6$ square degrees of spectroscopic overlap with the extended Groth Strip EGS DEEP2 survey \citep{2007ApJ...660L...1D,2013ApJS..208....5N}.  DEEP2 selects objects with $18.5 \leq R_{\rm AB} \leq 24.1$ with $\sim$60\% of objects with $i < 23$.  Finally, there are 23.1 square degrees of spectroscopic overlap with the VIMOS Public Extragalactic Redshift Survey \citep[VIPERS;][]{2014A&A...566A.108G} .  VIPERS selects $i<22.5$ galaxies with an additional colour selection to target galaxies in the redshift range $0.5 <z< 1.5$ and is thus highly incomplete at $z<0.5$.    We use redshift quality flags 3 and 4 ($>95$\% secure) for all of the spectroscopic surveys and create a matched catalogue in order to directly compare photometric redshifts with spectroscopic redshifts in Section~\ref{sec:bias} and Section~\ref{sec:bright}.

\subsection{The dependence of photometric redshift errors on galaxy type}
\label{sec:bias}
There is one main advantage of the methodology we propose in this paper over the reconstruction methods, discussed in Section~\ref{sec:intro}.  The unknown galaxy bias is fully accounted for in our modelling if the average galaxy bias of the population of outliers at a true redshift $z$ is not significantly different from the average galaxy bias of the main population at that same redshift.    If that is the case, we do not need to incorporate nuisance parameters that model scale-dependent and redshift-dependent galaxy bias in our analysis even though the typical photometric sample will have a different mean galaxy bias than the spectroscopic sample with which it is cross-correlated (e.g. the highly biased BOSS galaxies).  All these averaged galaxy bias properties are encompassed by the auto-correlation signal $w_{ii}^{\rm SP,O}$ (Eqn.~\ref{eq_crossterm}).  
Outliers in the photometric redshift measurements come from two sources.  The first is from random photometric errors as the majority of the faint galaxies used in cosmological analyses are detected below 10-$\sigma$. We would not expect this form of outlier to be galaxy-type dependent.  The second source, however, is template or training set degeneracies, where a low-redshift red galaxy has the same colour as a high-redshift intrinsically blue galaxy.   These outliers are clearly galaxy-type dependent.

\begin{figure}
  \includegraphics[width=0.485\textwidth]{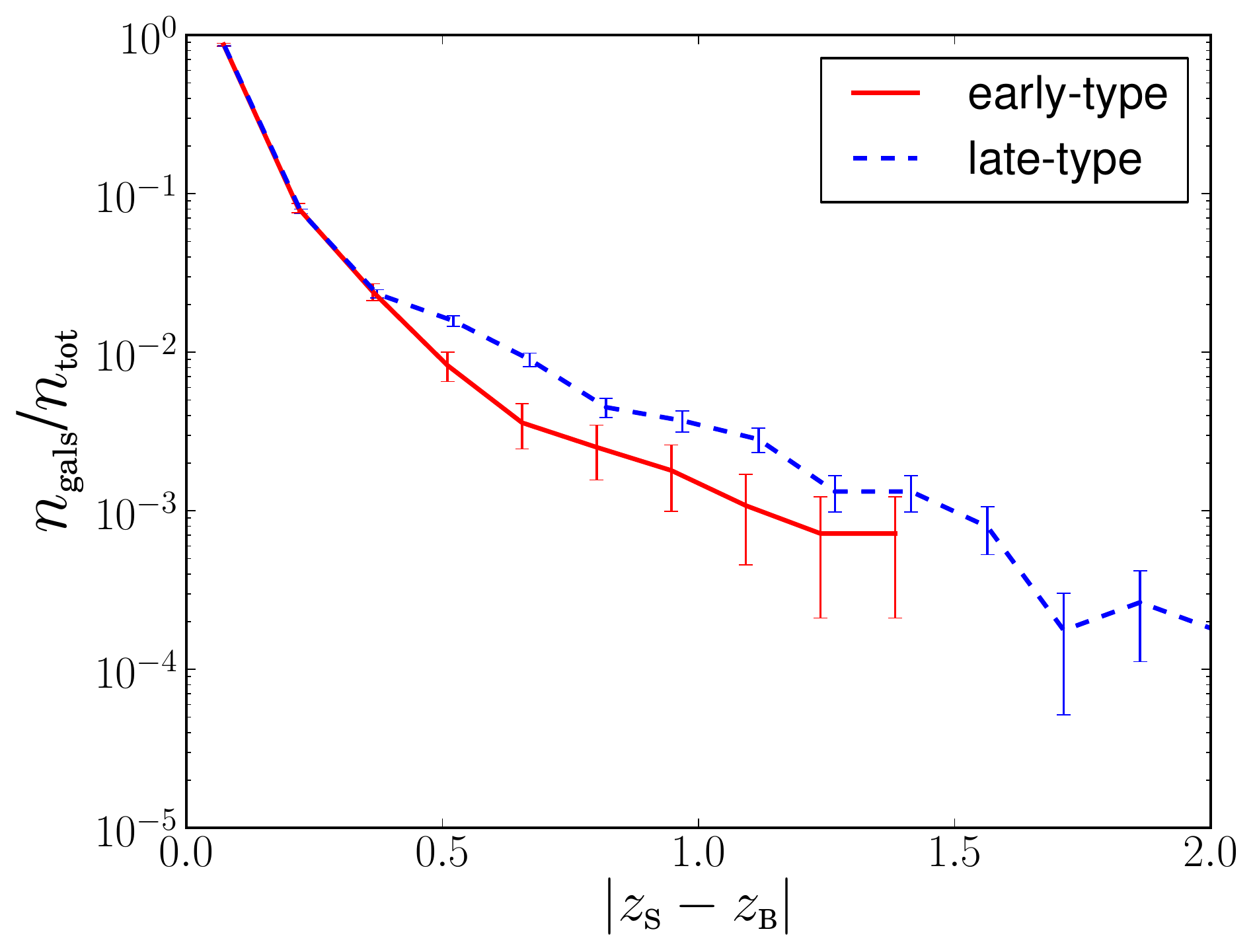}
  \caption{\label{fig:zs_minus_zp}
    The distribution of the absolute magnitude of the differential between
    spectroscopic redshift and photometric redshifts as a function of
    galaxy type as measured by \textsc{BPZ}. There are 2783 ET galaxies and 
11,341 LT galaxies.}
\end{figure}

To test the dependence of photometric redshift outliers on galaxy type, Figure~\ref{fig:zs_minus_zp} shows the absolute
difference between spectroscopic and photometric redshifts, $|z_{\rm s}-z_{\rm B}|$ for the CFHTLenS-VVDS matched catalogue described in Section~\ref{sec:VVD}.
Early-type galaxies are selected with the \textsc{BPZ} template type T$_{\rm B}<1.5$ (shown solid) 
and late-type galaxies are selected with $2<$T$_{\rm B}<4$ (shown dashed), 
where these ranges in T$_{\rm B}$ are shown by \citet{2014MNRAS.437.2111V} to
separate red and blue galaxies well.  These distributions have
been normalised by the total number of galaxies for the given type and are shown on a log-scale to enhance the differences.
We see that the numbers for early-type and late-type galaxies for $|z_{\rm s}-z_{\rm B}| > 0.4$ are different, with the majority of extreme outliers ($\Delta z > 0.2$) being late-type.   For the purposes of our analysis, the different galaxy bias properties of these extreme outliers would only impact upon the conclusions we drew from the most separated redshift bins where we find little signal to constrain the redshift distributions anyway.   
\label{sec:galbias}

\section{Results}
\label{sec:results}

\begin{figure*}
  \centering
  \includegraphics[width=\textwidth]{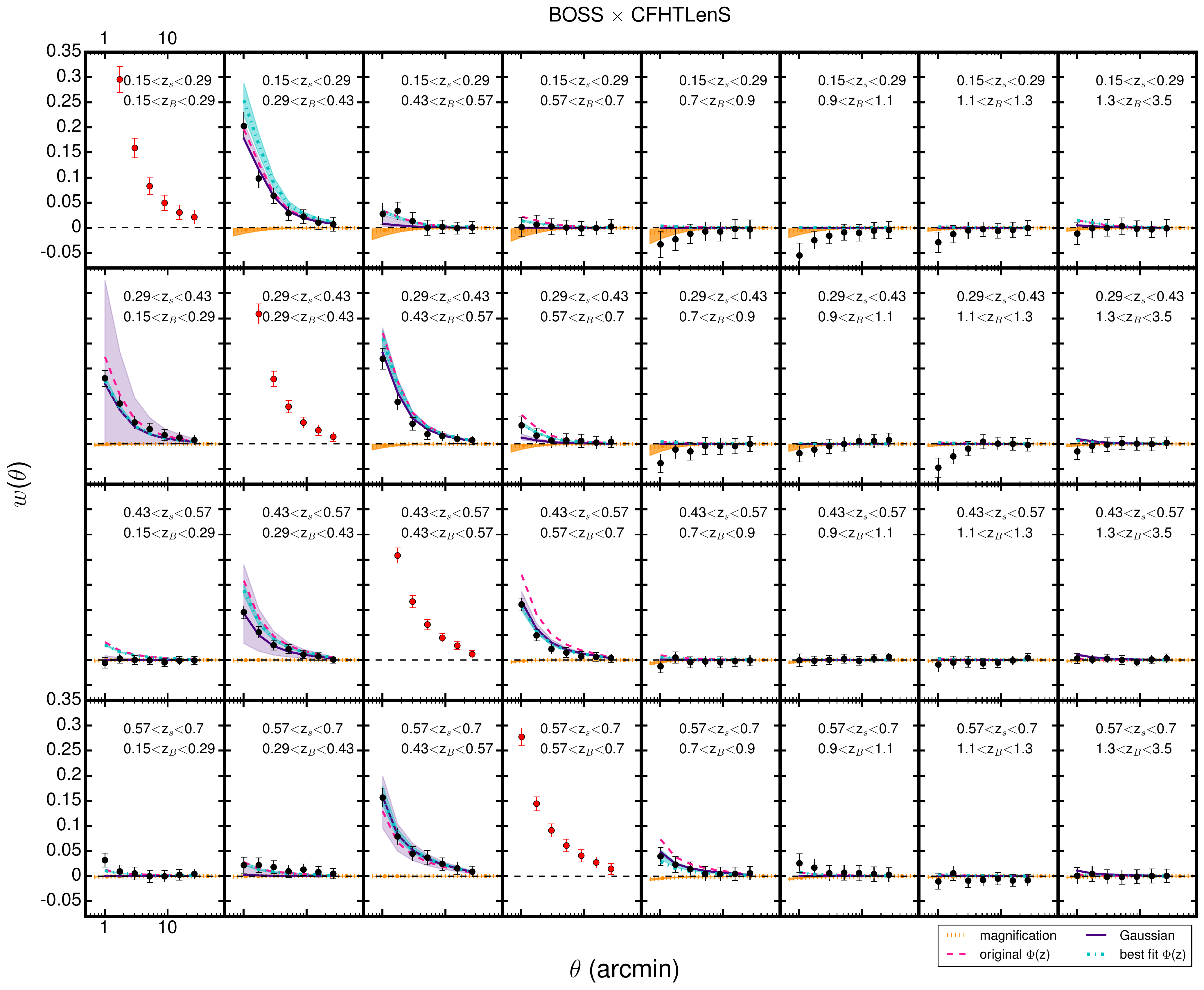}
  \caption{\label{fig:wij_cfhtboss}
    The measured cross-correlations between the CFHTLenS (photometric redshift) bin
    $j$ and
    BOSS (spectroscopic redshift) bin $i$ galaxies.  The data points
    are weighted means over the two CFHTLenS fields (W1 and W4), with
    the weights given by the number of pairs $(D_{i}D_{j})_{\theta}$ (see
    Eqn.~\ref{eq:LS}).  The red dashed line shows the
  predicted $ij$ clustering where $i \neq j$ based on Eqn.~\ref{eq_crossterm}.  The orange band shows
  a predicted magnification contribution based on halo model fits to the 
  BOSS galaxies from the literature.  The width of the orange band quantifies 
  how the predicted magnification signal changes for the uncertainties
  in the quoted halo 
  model parameters from the literature.  The cyan band shows the
  predicted $ij$ clustering after the $\Phi_{j}(z)$ have been shifted
  by the best fit quantities given in Table~\ref{tab:dzshift}.  The
  purple band shows the predicted $ij$ clustering given by best fit
  Gaussian $p(z)$.}
\end{figure*}

We measure the cross-correlations between three combinations of the 
photometric and spectroscopic surveys described above --  CFHTLenS-BOSS, RCSLenS-BOSS, and RCSLenS-WiggleZ. For the 8 redshift bins shown in
Figure~\ref{fig:wij_cfhtboss}, we measure the cross-correlation
$w_{ij}(\theta)$ using Eqn.~\ref{eq:LS} between spectroscopic redshift
bin $i$ and photometric redshift bin $j$ for 7 logarithmically-spaced angular bins in the range $1^{\prime} < \theta
< 35^{\prime}$.  Errors are jack-knife resampled such that each 1 deg$^{2}$ pointing is a
jack-knife sub-sample.  These sub-samples maximise the regularity in
the shapes of the jack-knife regions, as advocated by
\citet{2007MNRAS.381.1347C} and \citet{2009MNRAS.396...19N}.  This internal method of error estimation is approximate, as 
\citet{2009MNRAS.396...19N} have found jack-knife resampled covariance
matrices to be somewhat biased on small angular scales.  All of the
covariance matrices pass tests for positive definiteness and have
eigenvalues spanning a reasonable range.

\subsection{Photometric redshift accuracy in CFHTLenS}
In Figure~\ref{fig:wij_cfhtboss}, we present our results for the measured 
cross-correlations between the CFHTLenS and BOSS galaxies as filled 
circles.  Each panel represents a different cross-correlation between a spectroscopic and a photometric bin with the spectroscopic redshifts increasing from top to bottom and the photometric redshifts increasing from left to right.  The red dashed line shows the predicted clustering for the
off-diagonal panels using Eqn.~\ref{eq_crossterm}.  The orange solid band shows the predicted
magnification from Eqn.~\ref{eq:magcl} given HOD parameters taken from
\citet{2013MNRAS.429...98P} for the 
LOWZ sample and
\citet{2013arXiv1311.1480M} for the CMASS sample.  The $\alpha$ values used
are [0.52, 0.52, 0.46, 0.47, 0.45, 0.68, 0.89, 0.89] for CFHTLenS,
where each array value corresponds to the photometric redshift bins in
ascending order.  The errors obtained by
bootstrap resampling the $\alpha$ values are all on the order of $1
\times 10^{-3}$, so we do not explicitly quote them here.  The band of values is set 
by the minimum and maximum values possible
from sampling the HOD parameters within their 1-$\sigma$
uncertainties. The purple and cyan bands show the best fit models when
we allow for modification of the $\Phi_{j}(z)$ as described in
Section~\ref{sec:MCMCshift}.  We discuss these best fit models below in Section~\ref{improvedz}.

Focussing first on the data measurements, the diagonal cross-correlations are 
depicted in red and have the
highest amplitude, as is expected given the $\Phi_{j}(z)$ plotted in
Figure~\ref{fig:sumpdf}.  That is, the $\Phi_{j}(z)$ selected by $z_{B}$
peak are in the appropriate range.  However, the significant overlap between 
adjacent
bins also indicates that we would expect relatively high amplitudes in
the off-diagonal panels closest to the diagonal panels, and this trend
is confirmed in Figure~\ref{fig:wij_cfhtboss}.  The panels that
are furthest away from the diagonal reflect
cross-correlations with smaller amplitude.  Intrinsic clustering
can only cause positive cross-correlations, and thus, it is clear that
the most widely separated bins have little intrinsic clustering and a 
negligible amount of contamination between low-z photometric redshift bins and
high-z spectroscopic redshift bins and vice-versa.

Figure~\ref{fig:wij_cfhtboss} provides a wealth of information.  To
illustrate this, concentrate on a particular sample, the
$0.7<z_{B}<0.9$ bin.  Here, we find the leakage to the
$0.57<z_{s}<0.7$ bin as revealed by the non-zero cross-correlation
measured with the spectroscopic sample in this range, is well-modelled
by the best fit $\Phi_{j}(z)$.  The anti-correlation seen with lower spectroscopic redshift samples agrees with the predictions from lensing magnification.  Interestingly, some widely separated bins contain a negative signal (on the order
of -1\% at 1$^{\prime}$) in 1 or 2-$\sigma$ tension with the
magnification predictions.  These anti-correlations could potentially
be explained by systematic effects such as object detection and
deblending problems that are not fully characterised and accounted for
in the angular clustering measurements
\citep{2015MNRAS.449.1259S,2015arXiv150708336S,2015arXiv150904290M}.
In Appendix~\ref{sec:wthetasys} we investigate these systematic
effects and find that while they likely play a role in the negative
signals, we are unable to constrain their contribution without the aid of sophisticated image simulations.

Briefly returning to the question of galaxy bias discussed in Section~\ref{sec:bias} we repeat our analysis for the CFHTLenS-BOSS cross-correlation 
for the case of \textsc{BPZ} template type T$_{\rm B}<1.5$ and compare it to the
case of $2<$T$_{\rm B}<4$.   Here we find that $w_{ii}^{\rm SP,O}$ does differ in terms of amplitude and angular dependence 
for red and blue galaxies.  However, the angular dependence (i.e. shape) is similar over different photometric redshift bins,
which supports the assertion that the evolution of the scale-dependent photometric
galaxy bias is captured by the diagonal clustering measurements in each bin.
From this test and the analysis presented in Section~\ref{sec:bias} we conclude that varying galaxy bias does not impact upon the conclusions drawn in this paper.  However, the different galaxy biases in the outlier population will be important to model for reconstruction methods and future higher fidelity implementations of the methodology presented in this analysis.

\subsubsection{Improved CFHTLenS photometric redshift distributions}
\label{improvedz}
\begin{figure}
\includegraphics[width=0.48\textwidth]{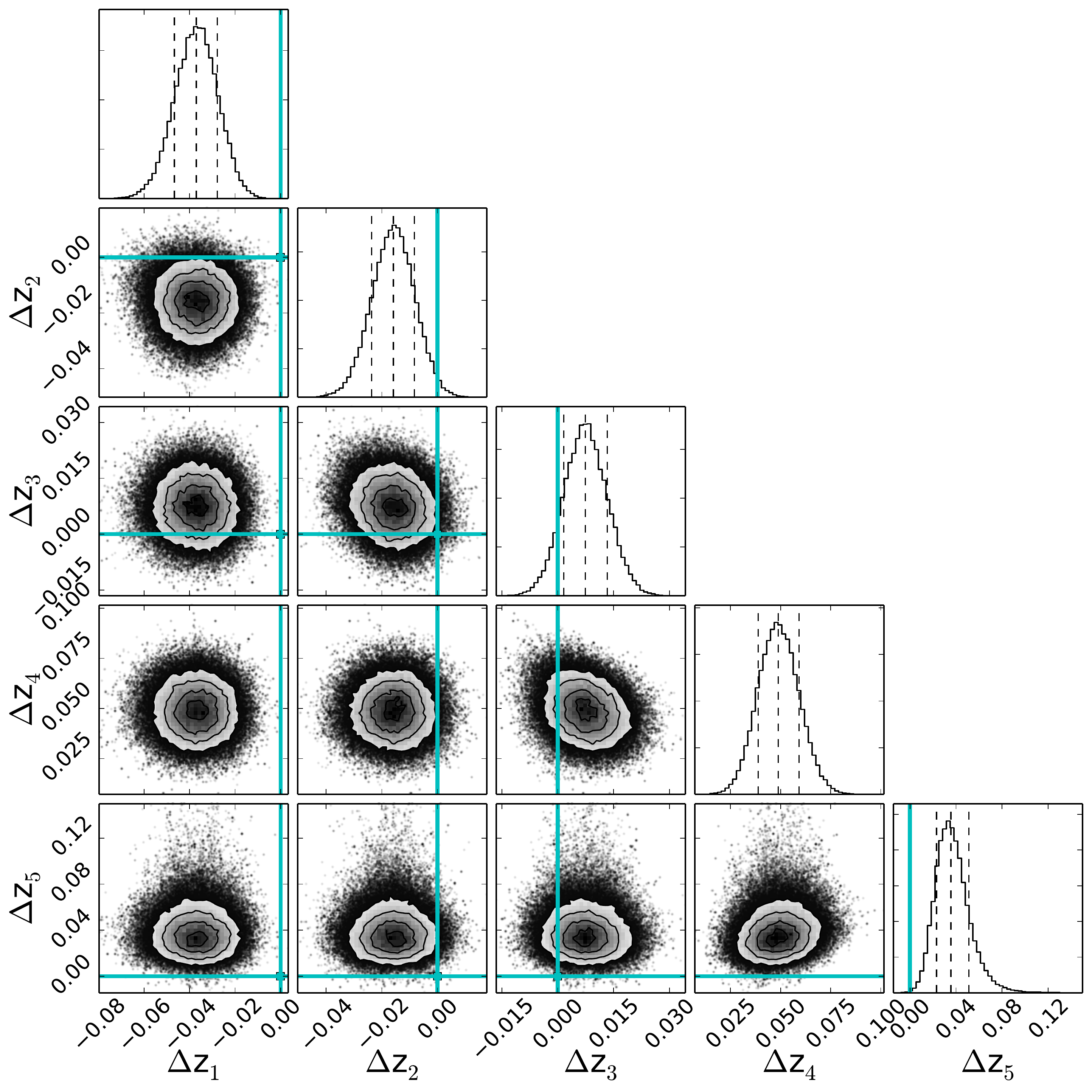}
\caption{\label{fig:amp_cfht_boss}
Joint fit shifts for photometric bin $j$ for CFHTLenS cross-correlated
with BOSS. The dashed black lines mark the 16th, 50th and 84th
percentiles of the samples in the marginalised distributions, and the
solid cyan lines mark $\Delta z_{j}=0$.}
\end{figure}

We determine best fit $\Phi_{j}(z)$ for each photometric redshift bin
using the procedure outlined in Section~\ref{sec:MCMCshift}, which
consists of two approaches.  The first method fits a shift to the
original \textsc{BPZ} $\Phi_{j}(z)$ along the $z$ dimension, and
Figure~\ref{fig:amp_cfht_boss} shows an example of the MCMC results
for CFHTLenS cross-correlated with BOSS.
Figure~\ref{fig:amp_cfht_boss} and subsequent corner plots were made
with \textsc{triangle.py}\footnote{https://github.com/dfm/corner.py} \citep{dan_foreman_mackey_2014_11020}. We impose the prior $-0.15<\Delta z_{j}<0.15$ to avoid sampling the parameter space where the $\Phi_{j}(z)$ swap positions.  The best fit values for the redshift bias in each redshift bin are given in Table~\ref{tab:dzshift}.
For the main results, we have chosen to only assign free parameters to the photometric
redshift bins that have at least one adjacent spectroscopic redshift
bin.  For CFHTLenS-BOSS, this limits us to the five photometric
redshift bins in the range $0.15<z_{B}<0.9$. In Figure D1 of
Appendix~\ref{sec:fullmcmc}, we show an example of the full parameter
sampling for all eight photometric redshift bins for CFHTLenS-BOSS,
where it is clear that there are degeneracies for the photometric
redshift bins that do not have an adjacent spectroscopic bin.  For
CFHTLenS-BOSS, we fit five free parameters for shifting the $\Phi_{j}(z)$, obtaining values of $\Delta z_{j}$ ranging from $-0.037^{+0.009}_{-0.010}$ for $j=1$ to $0.049^{+0.010}_{-0.010}$ for $j=4$.  This best fit is represented by the cyan band in Figure~\ref{fig:wij_cfhtboss}, where the width quantifies how the model changes for the uncertainties in the best fit model parameters.

The second method described in Section~\ref{sec:MCMCshift} is to fit Gaussian $\Phi_{j}(z)$, each with a mean, $\mu_{zj}$, and a standard deviation, $\sigma_{zj}$.  For CFHTLenS-BOSS, we fit ten free parameters
for Gaussian $\Phi_{j}(z)$.  We use the minimum and maximum redshifts of each bin $j$ defined in Section~\ref{sec:photdata}, $z_{\textrm{min},j}$ and $z_{\textrm{max},j}$, to impose the priors $z_{\textrm{min},j}<\mu_{zj}<z_{\textrm{max},j}$ and $0.001<\sigma_{zj}<z_{\textrm{max},j}-z_{\textrm{min},j}$.  These priors are chosen to avoid sampling the parameter space where the $\Phi_{j}(z)$ are extremely flat.  The values are provided in Table~\ref{tab:dzshift}.  We do not fit the
model to any data points corresponding to $z_{B}>0.9$ because the
outliers cannot be well-modelled by a single Gaussian.  In
Figure~\ref{fig:full_cfhtboss_gauss} of Appendix~\ref{sec:fullmcmc},
we show the MCMC sampling of the ten free Gaussian $\Phi_{j}(z)$
parameters.   Figure~\ref{fig:full_cfhtboss_gauss} shows much stronger
degeneracies than seen for the MCMC sampling of the shift method in Figure~\ref{fig:amp_cfht_boss}.  The best fit Gaussian model is represented by the purple band in Figure~\ref{fig:wij_cfhtboss}.  The large uncertainties are reflected in the width of the purple band.

We assess a goodness of fit using the reduced $\chi^{2}$, which is the
$\chi^{2}$ given by Eqn.~\ref{eq:chisq} divided by the degrees of
freedom (DOF).  The na{\"i}ve estimate of DOF is 387 = 392 data points
- 5 fit parameters for the shifted $\Phi_{j}(z)$ and 214 = 224 data
points - 10 fit parameters for the Gaussian $\Phi_{j}(z)$.  However,
see \citet{2010arXiv1012.3754A} for an argument that these DOF
estimates should be considered as upper limits, and a standard reduced
$\chi^2$ is not an appropriate measure of the goodness of fit.  As
previously mentioned, we also issue the caveat that our covariance
matrices are obtained via internal jack-knife resampling and may be somewhat biased.  Nonetheless, we quote the $\chi^2$/DOF to illustrate the performance of the fitting:  $\chi^{2}_{\rm no-shift}=508.06/387=1.31$ for the original \textsc{BPZ} $\Phi_{j}(z)$, $\chi^{2}_{\rm shift}=454.31/387=1.17$ for the \textsc{BPZ} $\Phi_{j}(z)$ after application of the best fit shifts, and $\chi^{2}_{\rm Gauss}=202.32/214=0.95$ for the best fit Gaussian $\Phi_{j}(z)$.  These values are also summarised in Table~\ref{tab:dzshift}.

\subsection{Comparison of photometric and spectroscopic redshifts for a bright CFHTLenS galaxy sample}
\label{sec:bright}
In this section, we compare the original and improved CFHTLenS redshift distributions from our method with those obtained directly by examining galaxies that have both spectroscopic and photometric redshifts.  For this comparison, we limit our photometric sample to bright galaxies with $i<23$ to match the completeness of the comparison spectroscopic sample described in Section~\ref{sec:VVD}.  We split the sample in four photometric redshift bins spanning $0.15<z_B<0.7$.  The total number of spectroscopic redshifts with matched photometric redshifts in the range $0.15<z_B<0.7$ are 3925 (VVDS), 3031 (EGS), and 11,108 (VIPERS). 

\begin{figure}
\includegraphics[width=0.485\textwidth]{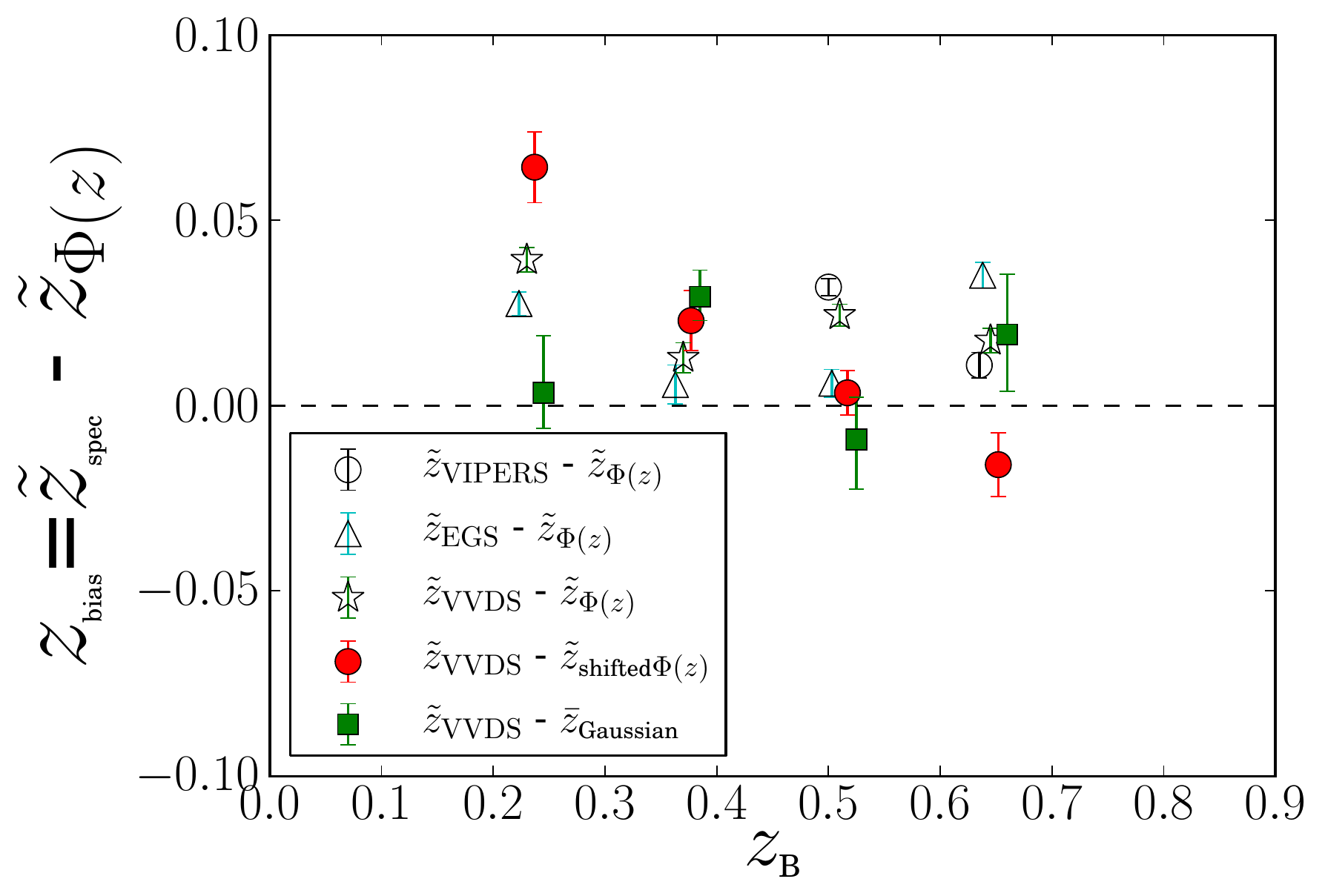}
\caption{\label{fig:speczcompbias}
  Direct comparison of the medians of the spectroscopic and photometric
  distributions for galaxies with $i<23$.  The open symbols show the difference between the medians
  of VIPERS (circles), DEEP2 EGS (triangles) or VVDS (stars) and the
  medians of the original \textsc{BPZ} $\Phi_{j}(z)$.  The closed
  symbols show the difference between the median of VVDS and the
  $\Phi_{j}(z)$ after application of the best fit shifts (circles) or
  the best fit Gaussian $\Phi_{j}(z)$ (squares).  The data points are
  slightly horizontally offset for clarity.  For the cross-correlation
  analysis, the CFHTLenS galaxies were cut to $i<23$ to be more consistent with the magnitude completeness limits of the comparison spectroscopic data sets.}
\end{figure}

For each photometric redshift bin, we define a measure of bias $z_{\rm
  bias}$ to be the difference between the median redshift
($\tilde{z}$) determined from the spectroscopic redshifts and from the
$\Phi_{j}(z)$.  We choose the median as it is less sensitive to incompleteness in the spectroscopic sample at $z_{\rm spec}>1.3$.  Error bars are determined by bootstrap resampling of the spectroscopic redshifts. 
Figure~\ref{fig:speczcompbias} shows $z_{\rm
  bias}$ determined from each of
the three spectroscopic surveys; VVDS (open star), VIPERS (open circle) and EGS
(open triangle).   Whilst these surveys are relatively complete we have not
investigated whether the sample is fully representative of the
photometric sample, both in terms of colour-space coverage \citep[see
for example][]{masters/etal:2015} and redshift coverage.  Even with
the applied bright magnitude limits there is a noticeable amount of
scatter even between the $z_{\rm bias}$ obtained from the two more
complete surveys VVDS and DEEP2 EGS, suggesting some sample variance
not taken into account by the bootstrap resampled error bars.  Residual
differences between the sampling of the galaxy populations can potentially be accounted for in future work using the re-weighting method of \citet{2015arXiv150705909B}.  For all bins, however, we conclude that the photometric redshifts underestimate the true median redshift of the galaxy sample.

We next measure the CFHTLenS-BOSS cross-correlations for bright
CFHTLenS galaxies with $i<23$ and determine redshift offsets and
best fit Gaussian distributions for the four redshift bins using our
MCMC analysis.  The resulting bias that we measure between the
spectroscopic redshift distribution and our improved photometric
redshift distribution is shown in Figure~\ref{fig:speczcompbias} where
we now use the median of the $\Phi_{j}(z)$ with best fit shifts
applied and the median of the best fit Gaussian, respectively.  We
find that applying the best fit shift to the photometric redshift
distributions leads to an even stronger underestimate of the true median redshift of
the galaxy sample (closed circles) for the lower two redshift bins.
However, the Gaussian model (closed squares) results in a $z_{\rm bias}$ that is
consistent with zero.  The total errors consist of the bootstrap resampled
errors from the spectroscopic redshifts added in quadrature with the
errors from the best fit values.  The latter could potentially be
slightly underestimated, as we ignore the covariance matrices between
the spectroscopic bins.

From this analysis we can conclude that modelling errors under the assumptions that the shape of the redshift distribution is accurate, and that the bias can be represented by a linear shift in the distribution is insufficient to capture the true underlying distribution when dealing with real data.   Our linear-shift model, also advocated by \citet{2015arXiv150705552T} and tested in Appendix~\ref{sec:valmocks}, was based on the well-modeled catastrophic outliers within CFHTLenS.  It does, however, overlook the potential for the width of the main peak in the distribution to vary.   For example, if the width were underestimated in the $\Phi_{j}(z)$, the model would underestimate the amplitude of the cross-correlation with both adjacent redshift bins.   A shift, in contrast, appears as an underestimate of the signal in one of the adjacent bins, with an overestimate in the other adjacent bin.  We see some evidence of this behaviour in the $i<23$ sample.

Our Gaussian $\Phi_{j}(z)$ model allows us to test the effects of being able to change the width of the redshift distribution as well as the mean, with the caveat that our single-moded Gaussian $\Phi_{j}(z)$ are too simple to characterise catastrophic outliers.   In this bright-galaxy measurement of bias in the main peak of the redshift distribution, the Gaussian model provides the most robust result.

\subsection{Photometric redshift accuracy in RCSLenS}
\begin{figure*}
  \includegraphics[width=\textwidth]{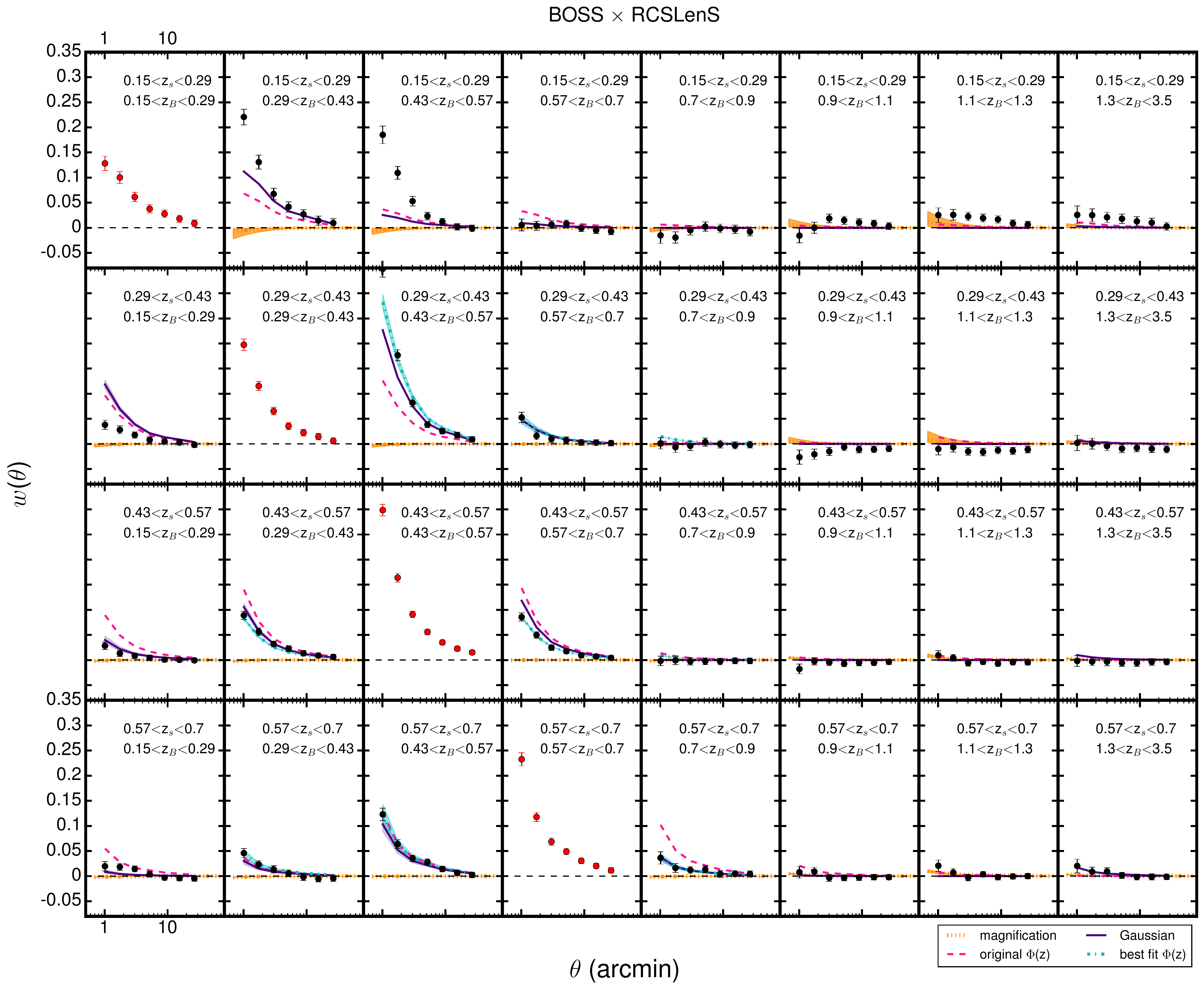}
  \caption{\label{fig:wij_rcsboss}
    As in Figure~\ref{fig:wij_cfhtboss}, cross-correlating RCSLenS
    (photometric redshift) bin $j$ and BOSS (spectroscopic redshift)
    bin $i$ galaxies. The data points
    are weighted means over the six RCSLenS fields that overlap BOSS, with
    the weights given by the number of pairs $(D_{i}D_{j})_{\theta}$ (see
    Eqn.~\ref{eq:LS}).}
  
\end{figure*}

\begin{figure*}
\includegraphics[width=\textwidth]{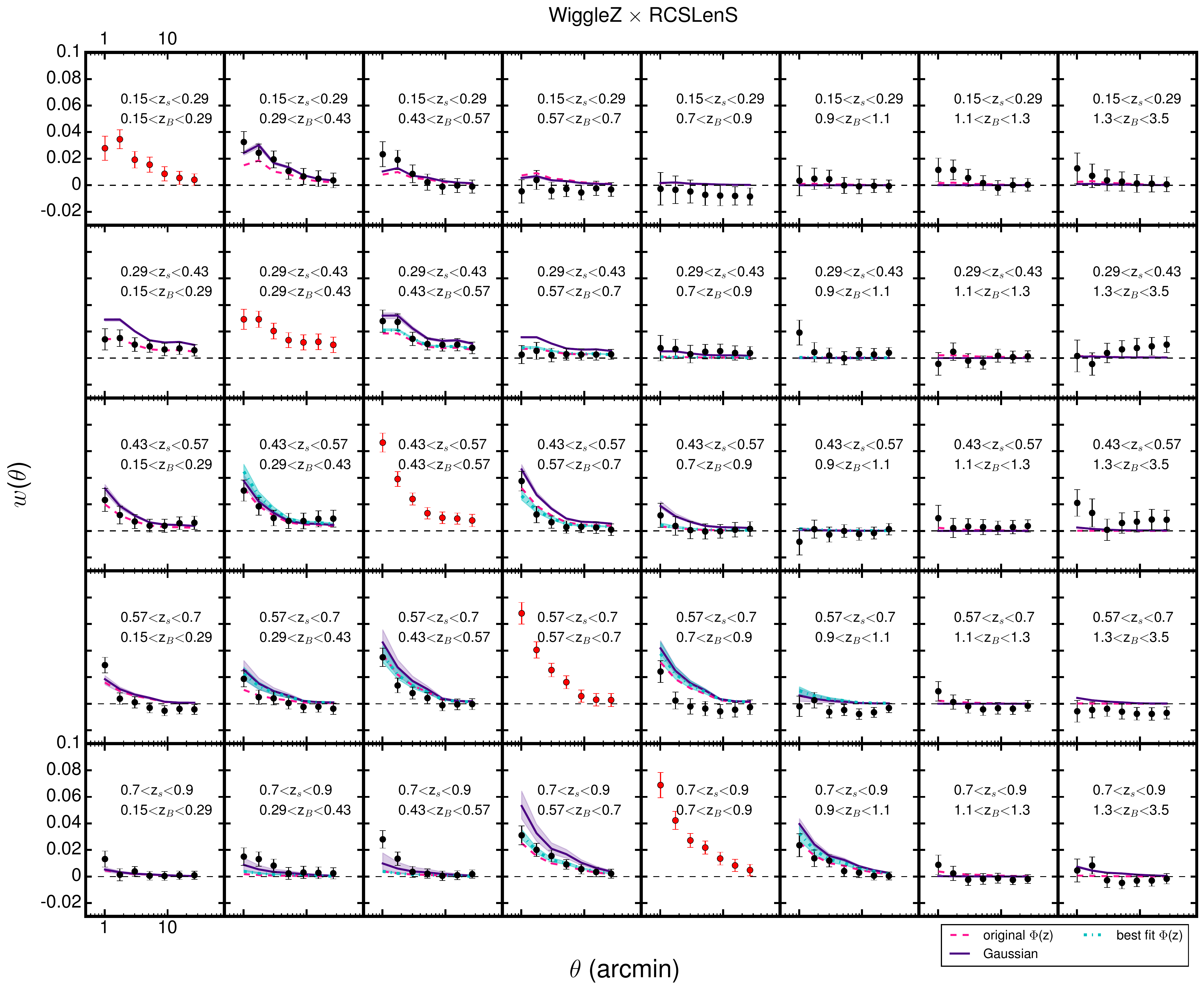}
  \caption{\label{fig:wij_rcswigglez}
     As in Figure~\ref{fig:wij_cfhtboss}, cross-correlating RCSLenS (photometric redshift) bin
    $j$ and
  WiggleZ (spectroscopic redshift) bin $i$ galaxies.  The data points
    are weighted means over the five RCSLenS fields that overlap WiggleZ, with
    the weights given by the number of pairs $(D_{i}D_{j})_{\theta}$ (see
    Eqn.~\ref{eq:LS}).  Note that there is no magnification prediction plotted here.}
\end{figure*}

In this section we repeat the cross-correlation analysis for the RCSLenS survey, but now with two spectroscopic surveys, BOSS 
(shown in Figure~\ref{fig:wij_rcsboss}) and WiggleZ (shown in Figure~\ref{fig:wij_rcswigglez}).
Comparing the measurements between RCSLenS and CFHTLenS (in Figure~\ref{fig:wij_cfhtboss}) we find the
amplitude and shape of the diagonal panels for the CFHTLenS and
RCSLenS cross-correlations with BOSS are similar; however, these are
qualitatively different from the RCSLenS cross-correlations with
WiggleZ.  This difference reflects the contrasting nature of the BOSS galaxy population
with galaxy biases given by $b_{\rm LOWZ}=1.6$ \citep{2013arXiv1312.4889C} and $b_{\rm
  CMASS}=1.9$ \citep{2014MNRAS.440.2692S} compared with the WiggleZ
galaxy population with $b_{\rm WiggleZ}=1.0$
\citep{2010MNRAS.406..803B}.  There are higher amplitudes of
correlation between the BOSS galaxies (with their higher mean galaxy bias) and both CFHTLenS and RCSLenS
galaxies compared to the lower amplitudes of correlation between the
WiggleZ galaxies (with their lower mean galaxy bias) and RCSLenS galaxies.  The typical signal to noise for RCSLenS-WiggleZ 
measurements is markedly less than for either the CFHTLenS-BOSS or the RCSLenS-BOSS cross-correlations.

In terms of the nature of the off-diagonal panels which reveal the strength of the photometric
redshift errors, there is a difference between the measurements for CFHTLenS-BOSS
and RCSLenS-BOSS highlighting the poorer quality of the 4-band RCSLenS photometric redshifts.
 The RCSLenS-BOSS off-diagonal measurements show positive signals, particularly in
the cross-correlations between the lowest spectroscopic redshift range
and the highest photometric redshift ranges (top right corner) and
between the lowest photometric ranges and all spectroscopic ranges
(leftmost column). This signal indicates the significant presence of
catastrophic outliers in the photometric redshifts for RCSLenS, which is
consistent with the characterisation of the photometric redshifts when
compared directly against spectroscopic redshifts in RCSLenS as
compared to CFHTLenS \citep{2016MNRAS.tmp.1132H}.

The orange band in Figure~\ref{fig:wij_rcsboss} is the magnification
signal computed in the same way as for Figure~\ref{fig:wij_cfhtboss}
with $\alpha$ values of [0.61, 0.42, 0.68, 0.94, 0.85, 1.29, 1.53,
1.13].  In contrast to CFHTLenS, the $\alpha$ values are greater than
one for the three highest redshift bins, leading to a positive
magnification signal as seen in the top right corner of
Figure~\ref{fig:wij_rcsboss}.  Focussing on $0.15<z_{s}<0.29$
cross-correlated with $1.1<z_{B}<1.3$, the orange band shows agreement
with the measured cross-correlations for the first two $\theta$ bins,
but the overall shape is qualitatively different.  This is likely
caused by an interplay between magnification and clustering of
catastrophic outliers;  the latter has not been accounted for in
the $\Phi_{j}(z)$ and thus is not reflected in our model.
There are no orange magnification prediction bands for
Figure~\ref{fig:wij_rcswigglez}, as
the WiggleZ galaxies are typically less massive than the BOSS
galaxies, and the amplitudes of their
magnification signals are likely correspondingly lower in the
  one-halo regime.  In the two-halo regime, only a couple of the
  cross-correlation measurements have a S/N over 2 (the
highest is 2.7), thereby obviating the need for magnification modelling.

\subsubsection{Improved RCSLenS photometric redshift distributions}
\label{sec:improvedrcs}

As in Section~\ref{improvedz}, we apply the methods of Section~\ref{sec:MCMCshift} to RCSLenS and summarise the best fit values for each survey combination in Table~\ref{tab:dzshift}.  We again assign free parameters to the photometric redshift bins that have at least one adjacent spectroscopic bin.  We impose a similar prior as in Section~\ref{improvedz}, except we extend the maximum possible shift to $0.3$ for the $j=4$ and $j=5$ bins.  For the shifting method, we additionally exclude the first redshift bin (the first row
and column of Figure~\ref{fig:wij_rcsboss}) from our analysis due to the high
catastrophic outlier rates indicated by the measurements.  The \textsc{BPZ} $\Phi_{j}(z)$ (and hence our model) do not account for these catastrophic
outlier rates and no amount of shifting will aid the cause. For RCSLenS-BOSS, we fit four free parameters for shifting the \textsc{BPZ} $\Phi_{j}(z)$ in the photometric redshift bins in the range $0.29<z_{B}<0.9$, obtaining values of $\Delta z_{j}$ ranging from $-0.095^{+0.007}_{-0.007}$ for $j=2$ to $0.236^{+0.026}_{-0.019}$ for $j=4$.  WiggleZ extends to slightly higher redshifts, thereby allowing
us to fit five free parameters for the photometric redshift bins from
0.29 to 1.1 for the case of RCSLenS-WiggleZ, obtaining values of $\Delta z_{j}$ ranging from $-0.040^{+0.029}_{-0.028}$ for $j=6$ to $0.070^{+0.019}_{-0.019}$ for $j=2$.  The best fit shift model is again represented by the cyan band for RCSLenS-BOSS in Figure~\ref{fig:wij_rcsboss} and for RCSLenS-WiggleZ in Figure~\ref{fig:wij_rcswigglez}.

We also fit a Gaussian $\Phi_{j}(z)$, each with a mean, $\mu_{zj}$, and a standard deviation, $\sigma_{zj}$.  As for CFHTLenS, we do not fit the model to any data points corresponding to $z_{B}>0.9$ for RCSLenS-BOSS and  $z_{B}>1.1$ for RCSLenS-WiggleZ because the outliers cannot be well-modelled by a single Gaussian.  We impose similar priors to those described in Section~\ref{improvedz}, but we extend the maximum possible $\sigma_{zj}$ by adding a value to account for photometric redshift scatter.  Specifically, we apply the prior $0.001<\sigma_{zj}<(z_{\textrm{max},j}-z_{\textrm{min},j})+0.08(1+z_{\textrm{mid},j})$ with $z_{\textrm{mid},j}$ being the midpoint between $z_{\textrm{min},j}$ and $z_{\textrm{max},j}$. For RCSLenS-BOSS, we fit ten free parameters.  For RCSLenS-WiggleZ, we fit twelve free parameters.  Both sets of best fit parameters are summarised in Table~\ref{tab:dzshift}.  The best fit Gaussian model is represented by the purple band for RCSLenS-BOSS in Figure~\ref{fig:wij_rcsboss} and for RCSLenS-WiggleZ in Figure~\ref{fig:wij_rcswigglez}.  The best fit Gaussian model also fails to reproduce the large signals seen in the cross-correlation between $0.15<z_{s}<0.29$ and the three photometric redshift bins in the range $0.15<z_{B}<0.57$ in Figure~\ref{fig:wij_rcsboss} (and to a lesser extent in Figure~\ref{fig:wij_rcswigglez}).

For the sake of comparison, we provide the $\chi^2$/DOF.  The DOF for RCSLenS-BOSS is 752 = 756 data points - 4 fit parameters for the shifted $\Phi_{j}(z)$ and 662 = 672 data points - 10 fit parameters for the Gaussian $\Phi_{j}(z)$.  The DOF for RCSLenS-WiggleZ is 835 = 840 data points - 5 fit parameters for the shifted $\Phi_{j}(z)$ and 863 = 875 data points - 12 fit parameters for the Gaussian $\Phi_{j}(z)$. The $\chi^2$/DOF for RCSLenS-BOSS are: $\chi^{2}_{\rm no-shift}=1280.44/752=1.70$ for the original \textsc{BPZ} $\Phi_{j}(z)$, $\chi^{2}_{\rm shift}=967.93/752=1.29$ for the \textsc{BPZ} $\Phi_{j}(z)$ after application of the best fit shifts, and $\chi^{2}_{\rm Gauss}=1155.4/662=1.75$ for the best fit Gaussian $\Phi_{j}(z)$.  The $\chi^2$/DOF for RCSLenS-WiggleZ are: $\chi^{2}_{\rm no-shift}=907.84/835=1.09$ for the original \textsc{BPZ} $\Phi_{j}(z)$, $\chi^{2}_{\rm shift}=891.59/835=1.07$ for the \textsc{BPZ} $\Phi_{j}(z)$ after application of the best fit shifts, and $\chi^{2}_{\rm Gauss}=823.94/863=0.95$ for the best fit Gaussian $\Phi_{j}(z)$.

\begin{figure}
\includegraphics[width=0.485\textwidth]{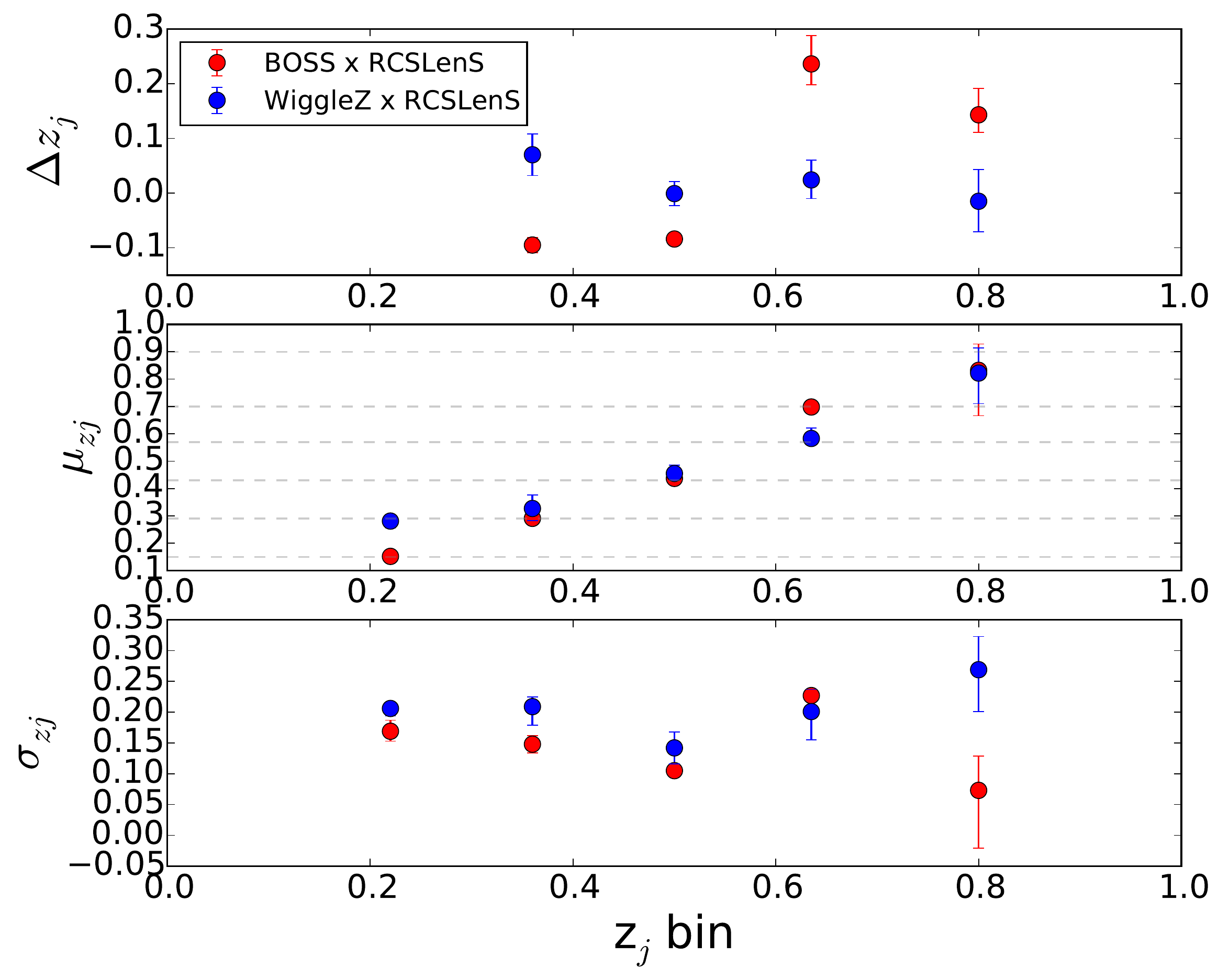}
\caption{\label{fig:comp_boss_wigglez}
Comparison of best fit parameters for RCSLenS from the
cross-correlation with BOSS and WiggleZ.  The top panel shows the best
fit shifts when the input models are the \textsc{BPZ} $\Phi_{j}(z)$.  The lower two
panels show the best fit means and
standard deviations when the input models are Gaussians.  The middle
panel contains dashed grey lines to indicate the boundaries of the
redshift bins.  The values and errors are summarised in
Table~\ref{tab:dzshift}.  Note that the plotted error bars correspond
to 2-$\sigma$, and there are no constraints for $\Delta z_{1}$, as explained in Section~\ref{sec:improvedrcs}.}
\end{figure}

Figure~\ref{fig:comp_boss_wigglez} shows the comparison of the best
fit $\Phi_{j}(z)$ shifts and best fit Gaussian $\mu_{zj}$ and
$\sigma_{zj}$.  Ideally, the best fit parameters from the BOSS
cross-correlation with RCSLenS would agree with those from the WiggleZ
cross-correlation with RCSLenS.   The best fit $\Phi_{j}(z)$ shifts
are significantly discrepant, whilst there is better agreement for the best fit Gaussian parameters.  We posit that the main source of this disagreement stems from our estimates of $\Phi_{j}(z)$ (from \textsc{BPZ} or Gaussians) being a poor characterisation of the underlying redshift distribution.  Future work might benefit from a more sophisticated model that retains the ability to model both the catastrophic outliers and the width of the redshift distribution.

\begin{table}
  \caption{\label{tab:dzshift} A summary of the best jointly fit
    $\Delta$z$_{j}$ shifts and best fit $\mu_{zj}$ and $\sigma_{zj}$
    for the three spectroscopic-photometric redshift survey
    combinations investigated in this work.  The effective unmasked
    overlap area is provided for each photometric/spectroscopic survey
    combination.  The reduced $\chi^{2}$ is also provided, with the
    sub-script `orig' for the \textsc{BPZ} $\Phi_{j}(z)$,
    `shift' for the \textsc{BPZ} $\Phi_{j}(z)$ after best fit shifts
    have been applied, and `Gauss' for the best fit Gaussians.}
  {\renewcommand{\arraystretch}{1.58}%
    \setlength{\tabcolsep}{0.01em}
    \begin{tabular}{lccc}
      \hline
      &CFHTLenS/BOSS~~&RCSLenS/BOSS~~&RCSLenS/WiggleZ\\
      \hline\hline
      $A_{\rm eff}$  &66.3  &180.2  &175.1  \\
      $\Delta z_{1}$  &$-0.037^{+0.009}_{-0.010}$  &-  &-  \\
      $\Delta z_{2}$  &$-0.016^{+0.008}_{-0.008}$   &$-0.095^{+0.007}_{-0.007}$  &$0.070^{+0.019}_{-0.019}$  \\
      $\Delta z_{3}$  &$0.007^{+0.006}_{-0.006}$  &$-0.084^{+0.005}_{-0.005}$  &$-0.001^{+0.011}_{-0.011}$  \\
      $\Delta z_{4}$  &$0.049^{+0.010}_{-0.010}$  &$0.236^{+0.026}_{-0.019}$  &$0.024^{+0.018}_{-0.022}$  \\
      $\Delta z_{5}$  &$0.036^{+0.016}_{-0.013}$  &$0.143^{+0.024}_{-0.016}$  &$-0.015^{+0.018}_{-0.022}$  \\
      $\Delta z_{6}$  &-  &-  &$-0.040^{+0.029}_{-0.028}$  \\
      $\chi^{2}_{\textrm{red,orig}}$ &1.31 &1.70 &1.09 \\
        $\chi^{2}_{\textrm{red,shift}}$ &1.17 &1.29 &1.07 \\
      $\mu_{z1}$  &$0.233^{+0.019}_{-0.020}$  &$0.152^{+0.003}_{-0.001}$  &$0.281^{+0.007}_{-0.013}$  \\
      $\mu_{z2}$  &$0.327^{+0.013}_{-0.017}$  &$0.291^{+0.002}_{-0.001}$  &$0.327^{+0.025}_{-0.022}$  \\
      $\mu_{z3}$  &$0.505^{+0.010}_{-0.012}$  &$0.437^{+0.005}_{-0.004}$  &$0.455^{+0.015}_{-0.013}$  \\
      $\mu_{z4}$  &$0.678^{+0.013}_{-0.014}$  &$0.698^{+0.002}_{-0.003}$  &$0.583^{+0.019}_{-0.010}$  \\
      $\mu_{z5}$  &$0.823^{+0.048}_{-0.056}$  &$0.832^{+0.048}_{-0.083}$  &$0.822^{+0.046}_{-0.056}$  \\
      $\mu_{z6}$  &-  &-  &$1.063^{+0.026}_{-0.041}$  \\
      $\sigma_{z1}$  &$0.052^{+0.023}_{-0.022}$  &$0.169^{+0.009}_{-0.008}$  &$0.206^{+0.003}_{-0.006}$  \\
      $\sigma_{z2}$  &$0.091^{+0.034}_{-0.017}$  &$0.148^{+0.007}_{-0.007}$  &$0.209^{+0.008}_{-0.015}$  \\
      $\sigma_{z3}$  &$0.098^{+0.026}_{-0.018}$  &$0.105^{+0.004}_{-0.004}$  &$0.142^{+0.013}_{-0.013}$  \\
      $\sigma_{z4}$  &$0.114^{+0.011}_{-0.016}$  &$0.227^{+0.002}_{-0.004}$  &$0.201^{+0.018}_{-0.023}$  \\
      $\sigma_{z5}$  &$0.0846^{+0.031}_{-0.038}$  &$0.073^{+0.028}_{-0.047}$  &$0.269^{+0.027}_{-0.034}$  \\
      $\sigma_{z6}$  &-  &-  &$0.188^{+0.056}_{-0.055}$  \\
      $\chi^{2}_{\textrm{red,Gauss}}$ &0.95 &1.75 &0.95 \\
      \hline
    \end{tabular}}
\end{table}

\section{Conclusions}
In this paper we have extended the formalism of
\citet{2010MNRAS.408.1168B} with the initial aim to verify galaxy
redshift distributions of a sample of galaxies, as determined from the
sum of their photometric redshift probability distributions,
$\Phi_{j}(z)$.   By cross-correlating the galaxy positions of
different photometric redshift bins, \citet{2013MNRAS.431.1547B}
showed that the measured galaxy clustering between photometric
redshift bins was consistent with the level of clustering expected
when the redshift distributions are estimated from \textsc{BPZ} in
this way.    For the cosmological analyses of CFHTLenS that then
followed
\citep{2013MNRAS.429.2249S,2013MNRAS.430.2200K,2013MNRAS.431.1547B,2013MNRAS.432.2433H,2014MNRAS.441.2725F,2014MNRAS.442.1326K},
the photometric error distribution was therefore assumed to be known
with zero uncertainty.  In this analysis we have increased
the fidelity of this test by using overlapping spectroscopy from BOSS
and, in the case of the RCSLenS, also from WiggleZ.  Cross-correlating
photometrically selected galaxy samples with galaxies binned by
spectroscopic redshift significantly enhances the signal to noise in the measured clustering between different redshift bins, thus allowing for a more stringent test of the photometrically derived redshift distributions.    For CFHTLenS, we can draw the same conclusions as \citet{2013MNRAS.431.1547B}, that the catastrophic outlier rate is well predicted by the $\Phi_{j}(z)$.  This can be seen by comparing the measured clustering signal with the model prediction, for widely separated bins ($\Delta z > 0.2$), in Figure~\ref{fig:wij_cfhtboss}.  Where the model is seen to fail, however, is around the peak of the redshift distribution, where significant deviations between the signal and model are seen for the first and fourth CFHTLenS redshift bin with $0.15<z_B<0.29$ and $0.57<z_B<0.7$.  A direct comparison between spectroscopic and photometric redshifts for a bright sample in Figure~\ref{fig:speczcompbias} also indicates that the $\Phi_{j}(z)$ model also fails in terms of the width or scatter in the redshift distribution.   This analysis implies that the cosmological analyses of CFHTLenS should have included systematic error terms in their analysis to account for bias and scatter in their redshift distributions that were not accurately modelled by the $\Phi_{j}(z)$.  

For RCSLenS, the conclusion that we can draw from this type of
analysis about the catastrophic outlier rate is complicated by the
fact that the lowest photometric redshift bin ($0.15<z_B<0.29$)
exhibits a strong catastrophic outlier rate when galaxy photometric
redshifts are individually compared to their directly measured
spectroscopic redshifts \citep{2016MNRAS.tmp.1132H}.  This outlier
rate is not predicted by the $\Phi_{j}(z)$ in this redshift range.
Model predictions for the cross-correlation of this low redshift bin
with higher photometric redshift bins will therefore be incorrect, as
they depend on the measured auto-correlation signal in this low
redshift bin (see Eqn.~\ref{eqn:Ewijmodel}).  A disagreement between
model and signal for cross-correlations with this bin will therefore
exist, even if the catastrophic outlier rate is accurately represented
in the higher redshift bins, as suggested by the direct comparison in
\citet{2016MNRAS.tmp.1132H}.  Excluding this low redshift bin from our analysis, we find that the catastrophic errors scattering galaxies from high redshift down to $z>0.29$ are well represented by the $\Phi_{j}(z)$.  Around the peak of each redshift distribution, however, significant deviations are again found, this time for the four bins spanning $0.29<z_B<0.9$.

As discussed in Section~\ref{sec:MCMCshift}, we can use our derived formalism to determine joint offsets in the peaks of the tomographic redshift distributions which are close in redshift to a spectroscopic sample.  To undertake this analysis we need to assume that the overall shape of each distribution is sufficiently accurate (i.e that the catastrophic outliers are well represented).   The offsets for each redshift bin that satisfy these constraints are given in Table~\ref{tab:dzshift}, showing significant biases up to $\Delta z \sim 0.236$.    We found that whilst the catastrophic outliers were well modelled by the $\Phi_{j}(z)$, the scatter was not, leading to inconsistencies when directly comparing our results with deep spectroscopic surveys in Section~\ref{sec:bright} and when comparing results between the BOSS and WiggleZ surveys.  
We therefore also determine the best fitting Gaussian $\Phi_{j}(z)$ for each bin which provides a more accurate estimate of the bias and scatter in each photometric redshift bin.  By definition however, this single-moded Gaussian is unable to model catastrophic outliers.

We have investigated the influence of the astrophysical features of galaxy bias and magnification, as well as the systematic effects of
object detection and deblending.  We found that these features do not
impact upon the conclusions presented here but will need to be 
investigated in more detail for future studies seeking to draw tighter constraints on
redshift distributions.  With the expected signal to noise of upcoming
deeper surveys, magnification may play a more significant role in the
angular cross-correlation signal between bins widely separated in
redshift.  Complete simulation pipelines including a full picture of
the underlying physics (clustering and lensing) and the observing, object detection and cataloguing process will be necessary to fully understand and disentangle the physical and systematic effects.

\subsection{Impact on cosmological parameter estimation with CFHTLenS}
Our methodology has been shown to provide a robust tool to verify redshift distributions for photometric surveys where overlapping but incomplete spectroscopy exists.  The recent existence of this ``same-sky'' survey data has allowed us to test the photometric redshift distributions used in the CFHTLenS weak lensing analyses with much higher fidelity than was previously possible.  
We can use the results of our analysis and scaling-relations from \citet{Jain_Seljak_1997} to estimate the impact on cosmological parameter estimation from using inaccurate redshift distributions in previous CFHTLenS analyses, for example \citet{2013MNRAS.432.2433H}.   

Weak lensing is most sensitive to a combination of the clustering amplitude $\sigma_8$ and the matter density parameter $\Omega_m$.  Defining $S_8 = \sigma_8 \Omega_m^{0.47}$, the two-point shear correlation function $\xi_+$ for a flat $\Lambda$CDM cosmology is related to $S_8$ as
\begin{equation}
\xi_+ \propto z_s^{1.52} S_8^{2.58} \,,
\end{equation}
for a single lensed source redshift slice at $z_s$ \citep{Jain_Seljak_1997} .   We can use this as a `toy-model' to indicate how errors in
the source redshift propagates into biases on cosmological parameters.
Considering the largest correction from Table~\ref{tab:dzshift}, a
bias of $0.049$ in the photometric redshift bin spanning
$0.57<z_B<0.7$ would correspond to an overestimate in the recovered
$S_8$ parameter from this tomographic bin by 4\%.   The second largest
correction from Table~\ref{tab:dzshift} is for the lowest redshift bin
which was already excluded from all CFHTLenS analyses as a result of
concerns over the photometric redshift accuracy in this bin
\citep{2012MNRAS.421.2355H}.  We refer the reader to \citet{2016arXiv160105786J} where the CFHTLenS tomographic cosmological analysis is revisited, taking into account the photometric redshift errors uncovered in this work.

\section*{Acknowledgments}

We are grateful to the RCS2 team for planning the survey, applying for
observing time, and conducting the observations. We acknowledge use of
the Canadian Astronomy Data Centre operated by the Dominion
Astrophysical Observatory for the National Research Council of
Canada’s Herzberg Institute of Astrophysics.  We thank Marcello
Cacciato and Maciek Bilicki for helpful comments on the manuscript and
Chris Morrison, Sarah Bridle and Marika Asgari for useful discussions.  
AC and CH acknowledge support from the European Research Council under
FP7 grant number 240185.  Part of this work was developed while at the
Aspen Center for Physics (NSF grant 1066293), and AC thanks the Center
for their hospitality.  CB acknowledges the support of the
Australian Research Council through the award of a Future Fellowship.
HH is supported by an Emmy Noether grant (No. Hi 1495/2-1)
of the Deutsche Forschungsgemeinschaft.  RN acknowledges support from the German Federal Ministry for Economic Affairs and Energy (BMWi) provided via DLR under project no.50QE1103.  MV acknowledges support from the European Research Council under FP7 grant number 279396 and the Netherlands Organisation for Scientific Research (NWO) through grants 614.001.103.

{\small \textit{Author Contributions:} All authors contributed to the
  development and writing of this paper.  The authorship list reflects the lead authors of this paper (AC, CH) followed by two
alphabetical groups. The first alphabetical group includes key
contributors to the science analysis and interpretation in this
paper.  CB made the angular clustering measurements, and HH led the RCSLenS collaboration.  
The second group consists of the founding core team and those whose significant efforts produced the final RCSLenS data product or data
products used in this analysis.
\bibliographystyle{mn2e}
\bibliography{sumpz_wtheta_bib}

\appendix

\section{Validation Tests on Mock Galaxies}
\label{sec:valmocks}

\begin{figure*}
 \includegraphics[width=\textwidth]{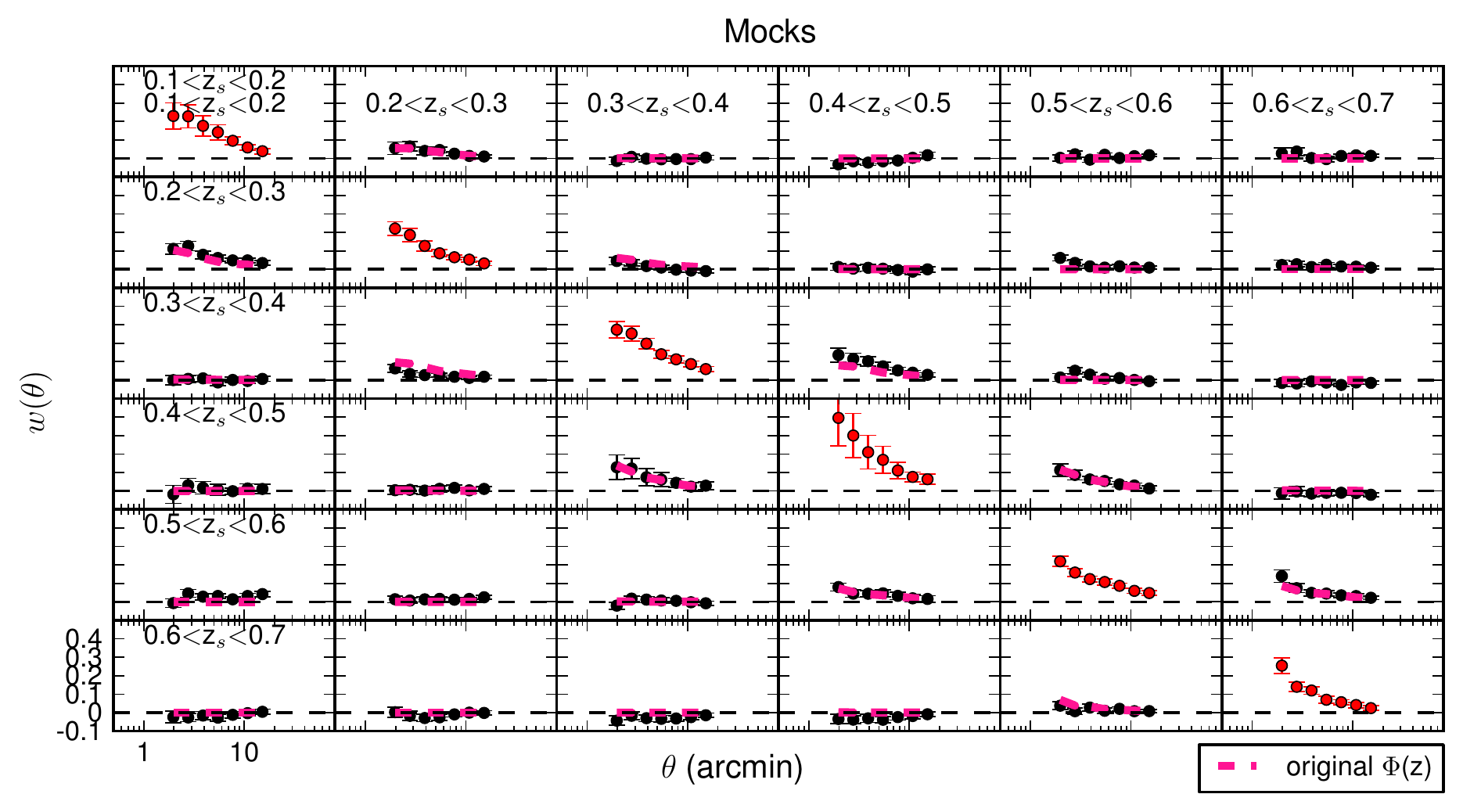}
  \caption{\label{fig:mockswtheta}
    Mock Analysis: The measured cross-correlations between the mock photometric galaxies and the mock spectroscopic galaxies.
}
\end{figure*}

We test the methodology described in Sections~\ref{sec:methodology}-\ref{sec:MCMCshift} on an idealised case consisting of lognormal distributions of 1,000,000 sources of known
clustering in a lightcone of area 25 deg$^{2}$, with an underlying Gaussian
$\Phi_{j}(z)$ with mean of 0.4 and a standard deviation of 0.3, cut in the range $0.1 < z < 0.7$. 10,000 sources are randomly picked as spectroscopic "BOSS" sources.  200,000 sources are randomly picked as photometric redshift "CFHTLenS"
sources.  The photometric redshift sources are assigned Gaussian scatters with mean of 0 and a standard
deviation of 0.05.   To simulate \textsc{BPZ}-like redshift probability
distributions, each source is assigned a $p(z)$, peaking at the scattered
photometric redshift value.  The spectroscopic and photometric sources are
divided into six redshift bins of width 0.1 in the range $ 0.1<z<0.7$, and
cross-correlations are measured.  We followed the steps outlined in
Section~\ref{sec:procedure} and show the measured cross-correlations
compared with the predicted cross-correlations in
Figure~\ref{fig:mockswtheta}. Figure~\ref{fig:mocksMCMC} shows the
results of MCMC sampling shifts of the redshift distributions for each
of the bins (described in Section~\ref{sec:MCMCshift}.  The best fit shifts are consistent with 0, as was input into the mocks, thus validating our method in this idealised scenario.

\begin{figure}
 \includegraphics[width=0.485\textwidth]{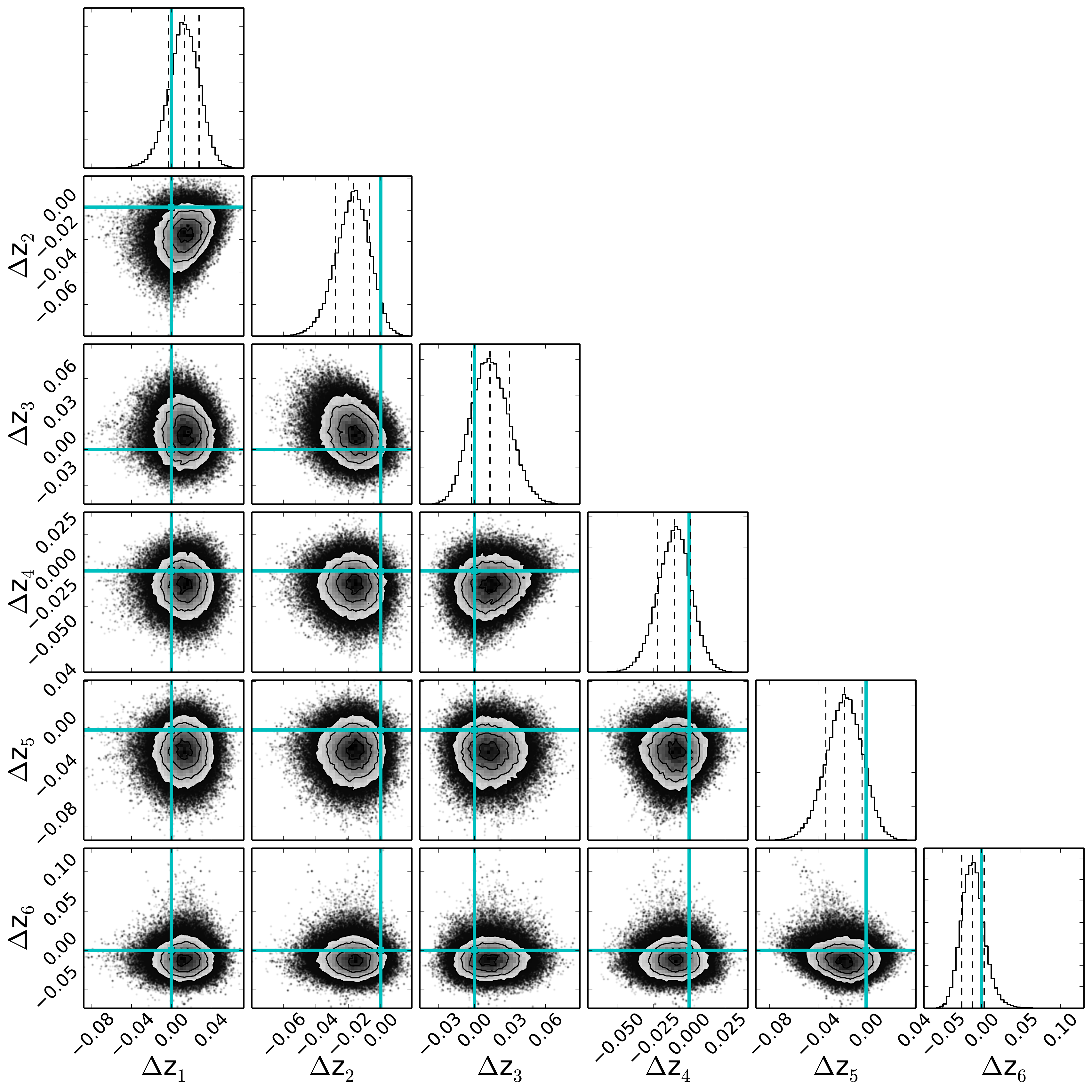}
  \caption{\label{fig:mocksMCMC}
    Mock Analysis: Best fit shifts in the redshift distributions for 6
    photometric redshift bins using joint MCMC sampling.
}
\end{figure}

\section{Halo Model}
\label{sec:halomodel}

The halo model provides an analytic framework for describing how
galaxies occupy dark matter halos, and the \textsc{chomp} software used in this
work follows the formalism of \citet{2000MNRAS.318..203S}.  We assume
cosmological parameters from \citet{2014A&A...571A..16P}, with
$\Omega_{\rm m}=0.315$, $\Omega_{\Lambda}=0.685$, $\sigma_{8}=0.829$,
$n_{\rm s}=0.9603$ and $\Omega_{\rm b}h^{2}=0.02205$.  Using the mass function from \citet{2002MNRAS.329...61S}, the density
profile from \citet{1996ApJ...462..563N}, and a halo bias model, we
can derive correlation functions.  To describe the numbers of central
and satellite galaxies as a function of halo mass, we assume the
functional forms given in \citet{2007ApJ...667..760Z}, using
 the best fit parameters from \citet{2013MNRAS.429...98P} for the LOWZ sample and
v1 of \citet{2013arXiv1311.1480M} for the CMASS sample into these functional
forms.  The number of central galaxies is given by
\begin{equation}
  N_{c}(M) = \frac{1}{2} \left [ 1+ \textrm{erf} \left ( \frac{\log(M)
      - \log(M_{\rm min})}{\sigma_{\log M}}
    \right ) \right ] \, ,
\end{equation}
where $M_{\rm min}$ is the minimum mass for a halo to host a central galaxy, and
$\sigma_{\log M}$ is the width of the cutoff.
The satellite term follows
\begin{equation}
N_{s}(M) = N_{c}(M) \left ( \frac{M-M_0}{M_{1}^{'}} \right )^{\alpha} \, ,
\end{equation}
where $M_{0}$ is the minimum mass for a halo to host satellite
galaxies, and $M_{1}^{'}$ is the mass differential at which a halo is
expected to have one satellite galaxy.  Adding the central and
satellite terms together gives the average
number of galaxies occupying a halo of mass M.

The LOWZ parameters taken from Table 3 of \citet{2013MNRAS.429...98P}
are $\log10{(M_{\rm min}/M_{\odot})}=13.25 \pm 0.26$,
  $\log10{(M_{1}^{'}/M_{\odot})}=14.18 \pm 0.39$, $\sigma_{\log M}=0.98
    \pm 0.57$, $\kappa=1.04 \pm 0.71$, and $\alpha=0.94 \pm 0.49$.
    Their $\kappa$ corresponds to our $M_{0}/M_{\rm min}$.  The CMASS parameters
    were originally taken from Table 2 of v1 of
    \citet{2013arXiv1311.1480M} as  $\log10{(M_{\rm min}/M_{\odot})}=13.21^{+0.13}_{-0.11}$,
  $\log10{(M_{1}^{'}/M_{\odot})}=14.15^{+0.09}_{-0.08}$, $\sigma_{\log
      M}=0.56^{+0.11}_{-0.09}$, $\kappa<0.58$, and $\alpha=1.06\pm
    0.49^{+0.11}_{-0.13}$.  The final published CMASS parameters
    appear in Table 1 of \citet{2015ApJ...806....2M} for three stellar
    mass subsamples.  The parameter values we use in this work fall
    within the range of values spanned by the three stellar mass
    subsamples in \citet{2015ApJ...806....2M}.

\section{Effects of Object Detection and Deblending on Clustering Measurements}\label{sec:wthetasys}

Clustering signals can contain not only contributions from physical phenomena 
like spatial correlation and magnification but also from systematic effects 
from the object detection and selection process itself. In this section, we first confirm that not accounting for small-scale ($<9''$) selection features in the random catalogues can affect the measured clustering on a range of scales;  we next investigate how object detection and deblending might affect the number of faint photometric galaxies in the vicinity of bright spectroscopic galaxies; finally, we perform a brute-force check by re-running our analysis on the data with photometric masks placed over the locations of known spectroscopic galaxies.

First, we set up a toy experiment to investigate how small scale
($<9.3''$) selection features in the random catalogues can affect the
measured clustering on a range of scales by creating three mock galaxy
catalogues based on the public N-body simulations described in
\citet{2012MNRAS.426.1262H}\footnote{http://www.cfhtlens.org}: a
`BOSS-like' clustered galaxy catalogue with $b_{\rm g}=2$ and a number
density of 1 arcmin$^{-2}$, a `photometric-like' clustered galaxy
catalogue with $b_{\rm g}=1$ and a number density of 10 arcmin$^{-2}$,
and a `no-clustering' galaxy catalogue with Poisson distribution and a number density
of 10 arcmin$^{-2}$.  For the photometric-like (PHOT) and
no-clustering (NC) galaxy
catalogues, we have full versions and masked versions where all of the
PHOT/NC galaxies within a $9.3''$ radius of a galaxy from
the BOSS-like catalogue are cut out.  The $9.3''$ radius is convenient because it
corresponds to 2 pixels in the original N-body simulation.  All
galaxies are at a single redshift $z=0.525$.  We measure the
$w(\theta)$ using Eqn.~\ref{eq:LS} for 6 different combinations quoted as [data
sample 2, random sample 2].  Data sample 1 is always the BOSS-like
catalogue, and random sample 1 is always a full (un-masked)
NC sub-sample. Where there are multiple sub-samples from
the NC galaxy catalogue used in the $w(\theta)$ measurement
(e.g. if both data sample 2 and random sample 2 are both drawn from
the NC galaxy catalogue), we ensure that the sub-samples
are mutually exclusive and do not share any of the same galaxies.  The combinations are:  1) [full PHOT, full NC], 2)
[masked PHOT, full
NC], 3) [masked PHOT, masked NC], 4) [full NC,
full NC], 5) [masked NC, full NC],
and 6) [masked NC, masked NC].  The results are
shown in Figure~\ref{fig:wthetatoy} where the combinations go from top
left to top right and continue from bottom left to bottom right.  The
left-most and right-most columns show the true cross-correlations when
the random catalogues properly account for the properties of the data
catalogues.  The top middle panel shows a dip at small $\theta$ in the cross-correlation
between a BOSS-like sample
and a masked PHOT sample with a corresponding
random sample that is \textit{not} masked.  Similarly, the lower
middle panel shows an anti-correlation for the $w(\theta)$ between a
BOSS-like sample and a masked NC sample with a corresponding random
sample that is not masked.

\begin{figure}
 \includegraphics[width=0.485\textwidth]{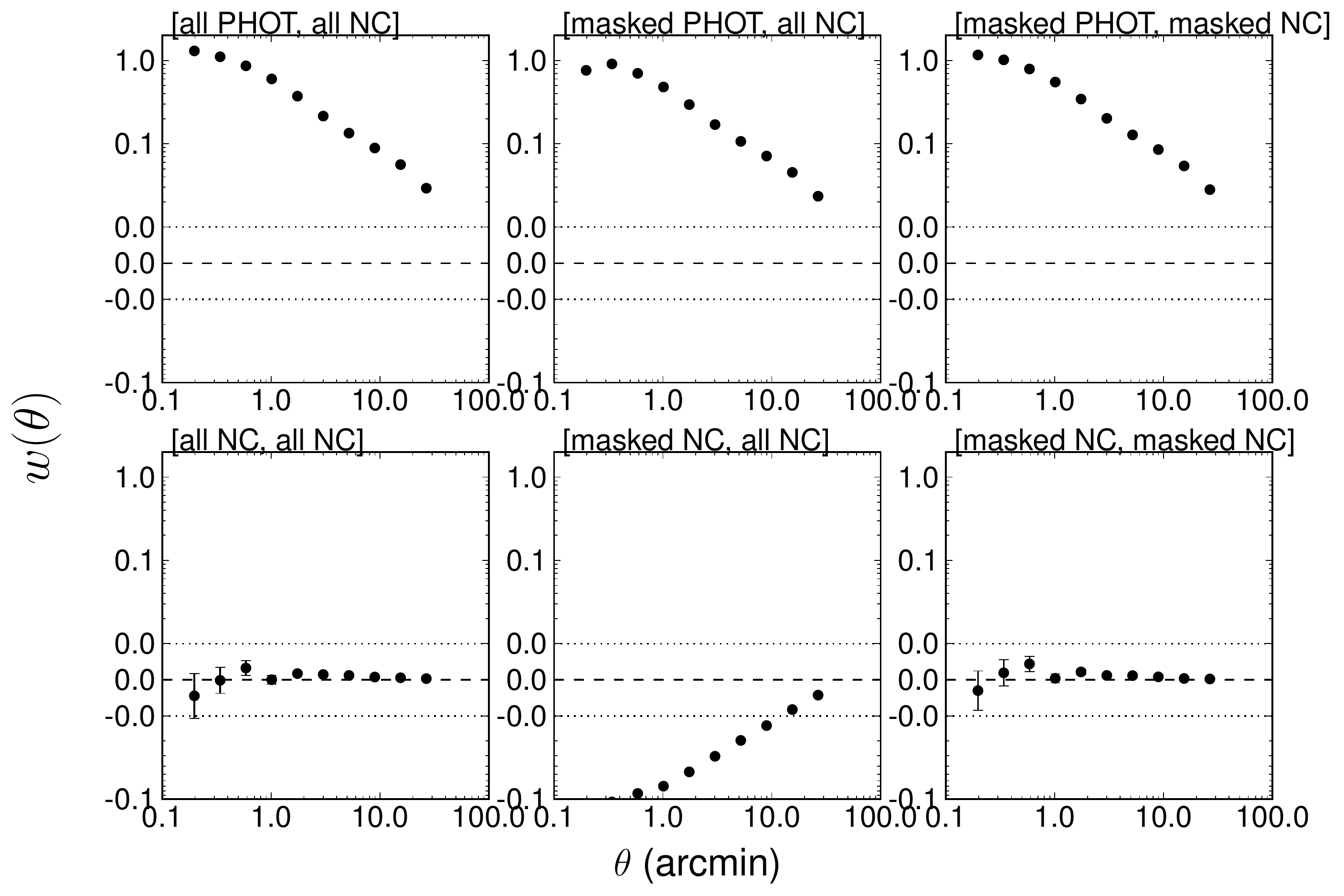}
  \caption{\label{fig:wthetatoy}
    Clustering measurements for different combinations of data and
    random mocks.  All combinations are measured with Eqn.~\ref{eq:LS}
    and involve a BOSS-like catalogue
    (data sample 1) and
    a corresponding random catalogue (random sample 1) that reflects the geometry of the
    BOSS-like catalogue.  The different combinations are labelled as
    [data sample 2, random sample 2], where the samples are drawn from
    either `masked' or `full' photometric-like (PHOT) or a
    no-clustering (NC) catalogues as described in the text.  The left
    and right columns of panels illustrate the case where the random
    sample properly accounts for the masking of the data sample.
    However, the middle column of panels shows that the signal is
    damped if masked galaxies
    in the data sample are not correspondingly masked in the random sample.
}
\end{figure}

The small-scale selection effects of object detection 
and deblending have been studied in the context of cluster galaxies which  
(typically bright, large in size, and residing 
in crowded fields) obscure nearby fainter or smaller galaxies 
\citep{2014MNRAS.439...48A,2014arXiv1405.4285M,2015MNRAS.449.1259S}.
The obscuration could, in principle, lead to a dearth of objects
detected close to the brighter galaxies and masquerade as a spurious
magnification signal.  Because the impact is uncertain, these
selection effects are not included in the CFHTLenS/RCSLenS masks.  We investigate possible systematic contributions to the clustering 
signal around BOSS galaxies from the cataloguing process by running 
simulations with \textsc{BALROG}\footnote{https://github.com/emhuff/Balrog}.
This public software allows us to add simulated galaxies to CFHTLenS images around known BOSS galaxies
and run SExtractor with the same object detection and deblending parameters used for
the actual catalogues.  We can repeat the process many times and measure
the recovered fraction of objects as a function of the angular
separation from the BOSS galaxies.  The results are shown in
Figure~\ref{fig:frac_rec_boss}, and a horizontal line marks the
average level
at which objects at any location in an image can be recovered (due to
noise). Figure~\ref{fig:frac_rec_boss} indicates that there is a lower fraction of objects recovered at angular scales smaller than $20''$.

\begin{figure}
 \includegraphics[width=0.485\textwidth]{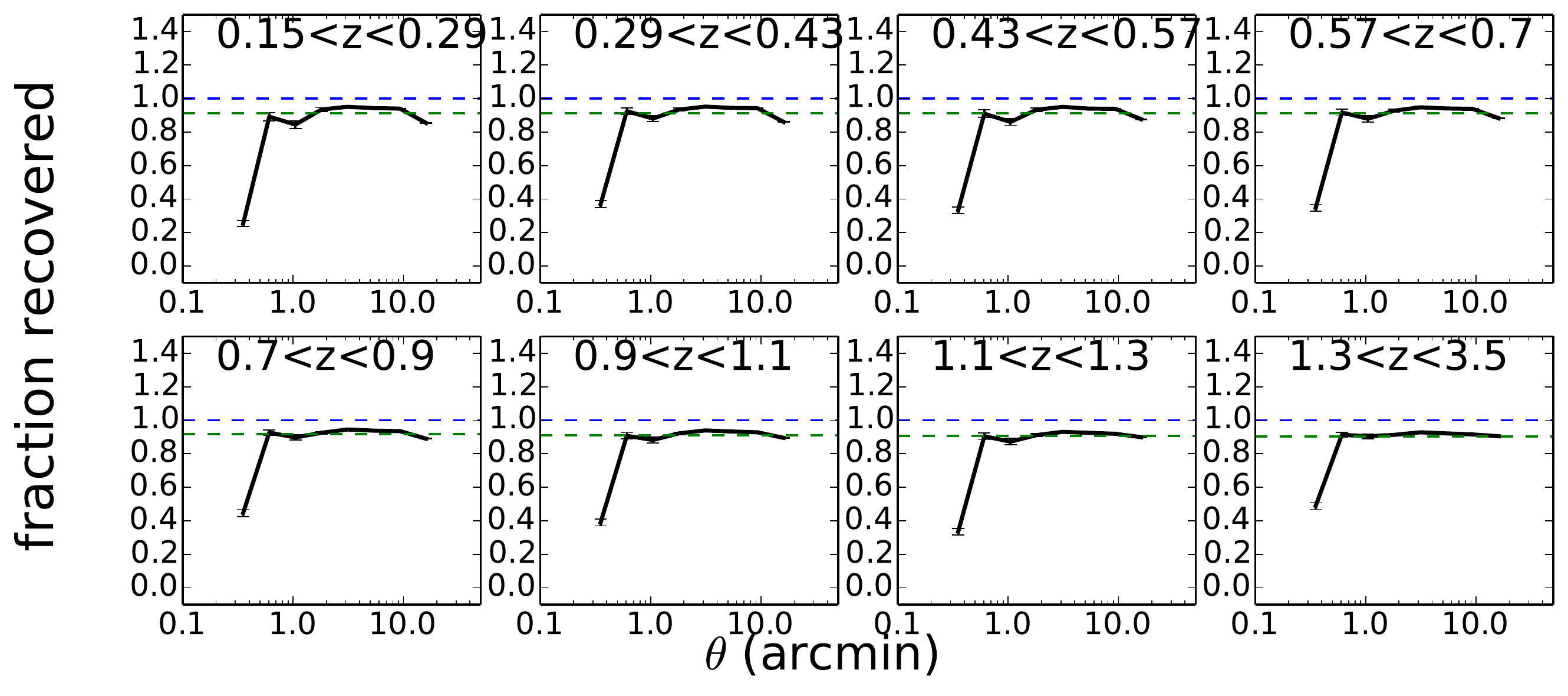}
  \caption{\label{fig:frac_rec_boss}
    Fraction of recovered objects from simulations.
}
\end{figure}

The conclusion we can draw from this analysis is that if the random
catalogue does not include small-scale selection effects, the measured
$w(\theta)$ will be diluted out to very large scales.  We re-run the
analysis with photometric catalogues that have $1'$ and $30''$ masks
centred on the BOSS galaxy positions in both the data and random
catalogues.  For both mask radii, the S/N renders the masked
$w(\theta)$ consistent with the $w(\theta)$ measured before masking.
Therefore, we conclude that this is not a dominant systematic in our
analysis but flag it as an important systematic in the future.  We leave further tests to future work. Much more sophisticated image simulations and mock galaxy catalogues will be required to fully disentangle physical effects (magnification) from systematic effects introduced in the observation and measurement process.

\section{Full MCMC For CFHTLenS-BOSS}
\label{sec:fullmcmc}

\begin{figure}
 \includegraphics[width=0.43\textwidth]{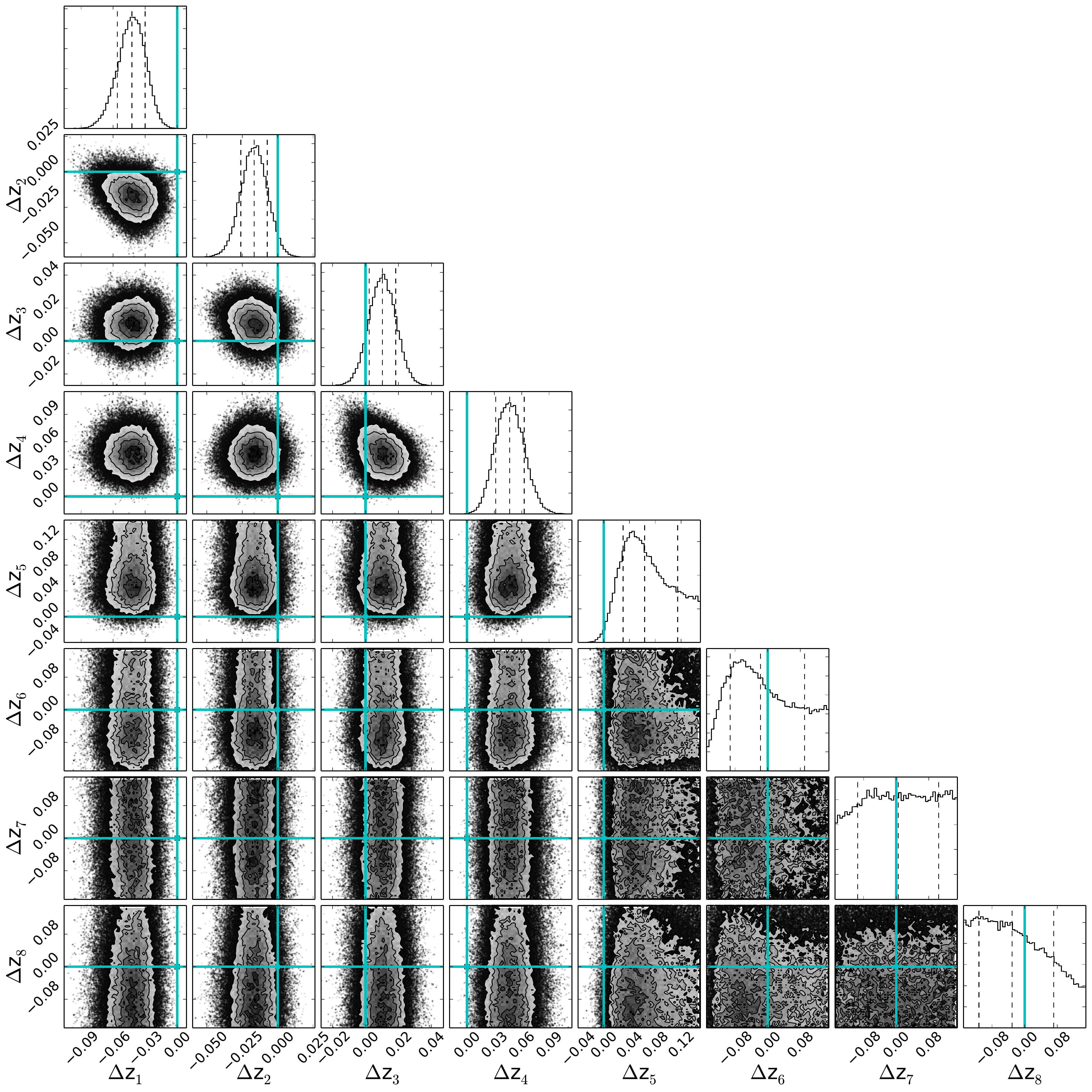}
  \caption{
    \label{fig:all_cfhtboss_mcmc} MCMC sampling of the redshift bias parameters for all 8 redshift bins for CFHTLenS-BOSS.
}
\end{figure}

In Figure~\ref{fig:all_cfhtboss_mcmc}, we show the results of MCMC sampling the redshift bias
parameters corresponding to all 8 redshift bins ([0.15, 0.29], [0.29,
0.43], [0.43, 0.57], [0.57, 0.7], [0.7, 0.9], [0.9, 1.1], [1.1, 1.3],
[1.3, 3.5]).  This example corresponds to the overlap between CFHTLenS
and BOSS.  As the spectroscopic coverage of BOSS only extends to the
first four redshift bins, there are clear degeneracies in the lower
half of Figure~\ref{fig:all_cfhtboss_mcmc} which reflect the fact that the spectroscopic redshifts up to z$\sim$0.7 can offer only limited information to constrain the higher redshift photometric galaxies.

In Figure~\ref{fig:full_cfhtboss_gauss}, we show the results of MCMC
sampling Gaussian $\Phi_{j}(z)$ described by their means $\mu_{zj}$ and
standard deviations $\sigma_{zj}$ for the same data set.

\begin{figure*}
  \includegraphics[width=\textwidth]{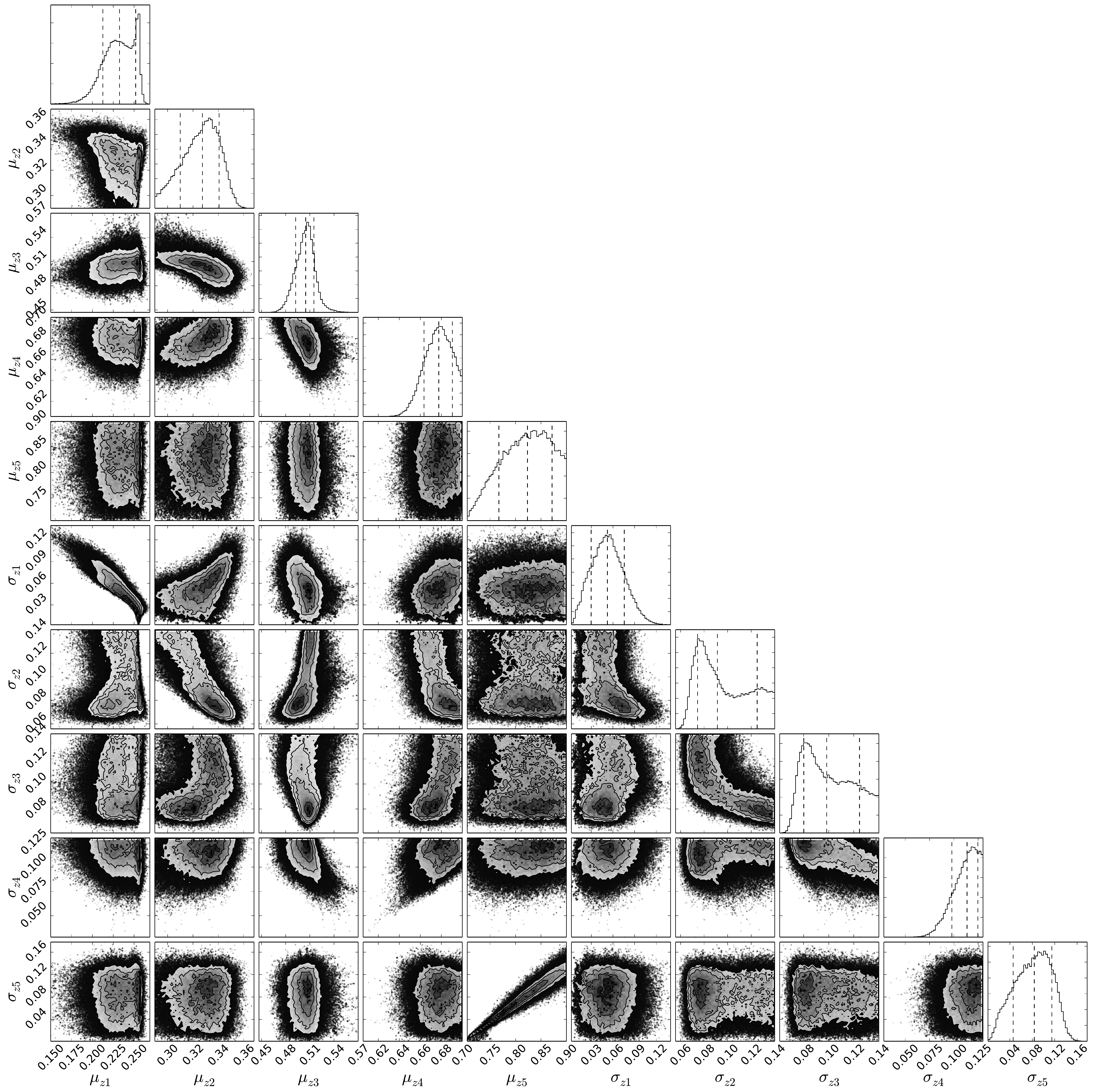}
  \caption{\label{fig:full_cfhtboss_gauss}
    MCMC sampling of the best fit Gaussian $\Phi_{J}(z)$ for 5 redshift bins
    for CFHTLenS-BOSS.  The redshift bin limits are given by [0.15, 0.29],
    [0.29, 0.43], [0.43, 0.57], [0.57, 0.7] and [0.7, 0.9].
}
\end{figure*}

\label{lastpage}

\end{document}